\documentclass[a4paper,fleqn,usenatbib]{mnras}

\usepackage{mathptmx}
\usepackage{graphicx}	% Including figure files
\usepackage{amsmath}	% Advanced maths commands
\usepackage{amssymb}	% Extra maths symbols

\def\teff{$T\rm_{eff}$}
\def\kms{$\mathrm{km\, s^{-1}}$}

\def\pris{Pristine\_221.8781+9.7844}

\newcommand{\mygi}{MyGIsFOS}

\title[The Pristine Survey IV]{The Pristine Survey IV: Approaching the Galactic metallicity floor with the discovery of an ultra metal-poor star\thanks{Based on observations collected at the European Organisation for Astronomical Research in the Southern Hemisphere under ESO programme 299.D-5042. Based on observations made with the William Herschel Telescope programme C31, operated on the island of La Palma by the Isaac Newton Group of Telescopes in the Spanish Observatorio del Roque de los Muchachos of the Instituto de Astrof'sica de Canarias. Based on observational programmes 15AC20, 15AF14, 15AF97, 16AC20, 16AC98, and 16AF14 obtained with MegaPrime/MegaCam, a joint project of CFHT and CEA/DAPNIA, at the Canada-France-Hawaii Telescope (CFHT).  }}
\author[E. Starkenburg et al.,]{Else Starkenburg$^1$\thanks{E-mail: estarkenburg@aip.de}, David S. Aguado$^{2,3}$, Piercarlo Bonifacio$^4$, Elisabetta Caffau$^4$, 
\newauthor Pascale Jablonka$^{4,5}$, Carmela Lardo$^{5}$, Nicolas Martin$^{6,7}$, Rub\'en S\'anchez-Janssen$^{8}$,  
\newauthor Federico Sestito$^{6}$, Kim A. Venn$^{9}$, Kris Youakim$^1$, Carlos Allende Prieto$^{2,3}$, Anke Arentsen$^1$, 
\newauthor Marc Gentile$^{5}$, Jonay I. Gonz\'{a}lez Hern\'{a}ndez$^{2,3}$, Collin Kielty$^{9}$, Helmer H. Koppelman$^{10}$, 
\newauthor Nicolas Longeard$^6$, Eline Tolstoy$^{9}$, Raymond G. Carlberg$^{11}$, Patrick C\^{o}t\'{e}$^{12}$, 
\newauthor Morgan Fouesneau$^7$, Vanessa Hill$^{13}$, Alan W. McConnachie$^{12}$, Julio F. Navarro$^{14}$ \\
$^1$ Leibniz Institute for Astrophyics Potsdam (AIP), An der Sternwarte 16, D-14482 Potsdam, Germany\\
$^2$ Instituto de Astrof\'{i}sica de Canarias, V\'{i}a L\'{a}ctea, 38205 La Laguna, Tenerife, Spain \\
$^3$ Universidad de La Laguna, Departamento de Astrof'sica, 38206 La Laguna, Tenerife, Spain \\
$^4$ GEPI, Observatoire de Paris, Universit\'e PSL, CNRS, Place Jules Janssen, 92190, Meudon, France \\
$^5$ Institute of Physics, Laboratoire d'astrophysique, \'{E}cole Polytechnique F\'{e}d\'{e}rale de Lausanne (EPFL), Observatoire, 1290 Versoix, Switzerland\\
$^6$ Universit\'e de Strasbourg, CNRS, Observatoire astronomique de Strasbourg, UMR 7550, F-67000 Strasbourg, France\\
$^7$ Max-Planck-Institut f\"{u}r Astronomie, K\"{o}nigstuhl 17, D-69117 Heidelberg, Germany \\
$^{8}$ UK Astronomy Technology Centre, Royal Observatory Edinburgh, Blackford Hill, Edinburgh, EH9 3HJ, UK \\
$^9$ Dept. of Physics and Astronomy, University of Victoria, P.O. Box 3055, STN CSC, Victoria BC V8W 3P6, Canada \\
$^{10}$ Kapteyn Astronomical Institute, University of Groningen, Landleven 12, 9747AD Groningen, Netherlands\\
$^{11}$ Department of Astronomy \& Astrophysics, University of Toronto, Toronto, ON M5S 3H4, Canada \\
$^{12}$ NRC Herzberg Astronomy and Astrophysics, 5071 West Saanich Road, Victoria, BC V9E 2E7, Canada \\
$^{13}$ Laboratoire Lagrange, Universit\'e de Nice Sophia-Antipolis, Observatoire de la C\^{o}te d'Azur, CNRS, Bd de l'Observatoire, \\
CS 34229, 06304 Nice cedex 4, France \\
$^{14}$ Dept. of Physics and Astronomy, University of Victoria, P.O. Box 3055, STN CSC, Victoria BC V8W 3P6, Canada \\
}

\date{Accepted XXX. Received YYY; in original form ZZZ}

\pubyear{2018}

\begin{document}
\label{firstpage}
\pagerange{\pageref{firstpage}--\pageref{lastpage}}
\maketitle

\begin{abstract}
The early Universe presented a star formation environment that was almost devoid of heavy elements. The lowest metallicity stars thus provide a unique window into the earliest Galactic stages, but are exceedingly rare and difficult to find. Here we present the discovery of an ultra-metal-poor star, Pristine\_221.8781+9.7844, using narrow-band Ca H\&K photometry from the \textit{Pristine} survey. Follow-up medium and high-resolution spectroscopy confirms the ultra-metal-poor nature of \pris\  ([Fe/H] = $-$4.66 $\pm$ 0.13 in 1D LTE) with an enhancement of 0.3$-$0.4 dex in $\alpha$-elements relative to Fe, and an unusually low carbon abundance. We derive an upper limit of A(C)  = 5.6, well below typical A(C) values for such ultra metal-poor stars. This makes \pris\  one of the most metal-poor stars; in fact, it is very similar to the most metal-poor star known (SDSS J102915+172927). The existence of a class of ultra metal-poor stars with low(er) carbon abundances suggest that there must have been several formation channels in the early Universe through which long-lived, low-mass stars were formed.
\end{abstract}


%%%%%%%%%%%%%%%%%%%%%%%%%%%%%%%%%%%%%%%%%%%%%%%%%%

%%%%%%%%%%%%%%%%% BODY OF PAPER %%%%%%%%%%%%%%%%%%

\section{Introduction}

The search for the most metal-poor stars has been compared to finding a needle in a haystack. In a typical halo field,
only one in $\sim$80,000 stars is expected to have [Fe/H]$< -4$ \citep{PristineIII}. Although much effort has been
devoted to the discovery and study of {\it extremely}, {\it ultra}, and {\it hyper} metal-poor
stars with [Fe/H] $< -3.0$, [Fe/H] $< -4.0$, and [Fe/H]$ < -5.0$, respectively,\footnote{See \citet{Beers05} for details on these definitions.} their overall numbers remain small. Still, these stars paint a fascinating picture about chemical enrichment in the very early Galaxy and the physics of star formation in environments that were mostly devoid of metals.

The first stars that formed in the Universe necessarily contained only hydrogen, helium and traces of lithium.
However, no star with such a primordial composition has been observed to date. As such, it is heavily debated whether long-lived stars (of lower mass) were formed in this epoch. Theoretical studies point out that, in gas environments devoid of heavier 
elements, cooling is more problematic and therefore proto-stars will be heavier and shorter lived. It is unclear, though, if in the fragmentation 
of the proto-stellar cloud any stars with masses lower than one solar mass could be formed that would have lifetimes similar to the
age of the Universe \citep[see][and references therein]{Bromm13,Greif15}.

In addition to cooling through metallic atomic lines, it is thought that dust grains can be an important cooling 
mechanism in very  metal-poor environments, bringing down the critical metallicity to allow cooling in lower mass proto-stellar 
clouds \citep[see e.g.,][]{Omukai08, Schneider12a, Schneider12b, Chiaki17}.  

There are presently 12 stars known to have intrinsic iron-abundances below [Fe/H] = $-$4.5 \citep{Christlieb02,Frebel05, Norris07,Caffau11,Norris12,Keller14, Hansen14, Allende15, Frebel15, Bonifacio15,Caffau16, Bonifacio18a, Aguado18a, Aguado18b}. The SkyMapper Southern Sky Survey star SMSS J031300.36-670839.3, which is the current record holder amongst iron-poor stars at [Fe/H] $< -7.0$, shows a very high carbon abundance of [C/Fe]$ > 5$ \citep{Nordlander17}. Based on the handful of stars found in the ultra metal-poor regime, this seems very typical for this metallicity regime; almost all ultra iron-poor stars show a very high carbon abundance  \citep[see e.g., the compilations in][]{Norris13,Frebel15,Aguado17}. It has been noted that there seems to be a trend with increasing carbon-to-iron ratio as [Fe/H] decreases, and for many of the most iron-poor stars, the absolute abundance of carbon lies around a value of A(C) $\sim$ 6.5 \citep{Spite13,Yoon16}\footnote{In this notation A(X) = log ($N_{\rm{X}}$/$N_{\rm{H}}$)+12, where X represents a given element.}. While the main focus in classification diagnostics has been on the (more readily measureable) carbon abundance,  other light elements, such as nitrogen, oxygen, and sodium, are also often greatly enhanced in these stars with respect to solar [X/Fe] abundance ratios. However, there are certainly stars observed that do not follow this trend. The most \textit{metal-poor} star known today is the ultra metal-poor star SDSS J102915+172927, that was shown not to be highly carbon-enhanced, but instead to have A(C) $<$ 4.2 at [Fe/H] $\sim$ --5.0. This star, as well as some other stars with only mild carbon enhancements \citep[e.g.,][]{Norris12}, suggest that there might be multiple formation routes for ultra metal-poor stars, with important consequences for theories of early star formation in the Galaxy. On the other hand, very recently a new hyper iron-poor star, SDSS J0815+4729, has been discovered showing a extremely high carbon abundance A(C)~$\sim 7.7$~dex \citep{Aguado18a}. 

We report here the discovery of Pristine\_221.8781+9.7844, an ultra low-metallicity star which belongs to the rare class of objects that have both low [Fe/H] as well as [C/Fe] abundances. This star was discovered thanks to the discriminatory power of narrow-band Ca H\&K photometry from the \textit{Pristine} survey \citep[see][]{PristineI, PristineII, PristineIII}, in combination with broad-band photometry from the Sloan Digital Sky Survey \citep[SDSS,][]{SDSS16}. The effectiveness of such narrow-band imaging techniques --- or, alternatively, very low-resolution prism spectroscopy in this same wavelength region --- has been convincingly demonstrated in the past \citep[e.g.,][]{Beers85,Anthony91,Anthony00, Christlieb02,Keller07,Murphy09,Howes15,Koch16}. The discovery of this new star demonstrates the increasing efficiency with which astronomers are now able to identify ``the needles in the haystack''. 

In Section \ref{sec:phot}, we describe the initial selection of this star and the derivation of stellar parameters from photometry and astrometry. In Section \ref{sec:mr} we describe the medium-resolution spectrum and its analysis, and in Section \ref{sec:hr} the high-resolution spectrum used to study its chemical properties. We calculate the abundances from absorption features in Section \ref{sec:abu}, and discuss the uncertainties on these measurements due to uncertainties in stellar parameters as well as 3D non-LTE effects. Finally, in Section \ref{sec:compare}, we show how \pris\ compares with other stars having similar metallicities.

\section{Photometry}\label{sec:phot}

\subsection{Photometric selection}
Pristine\_221.8781+9.7844 was first identified in our
narrow-band photometric survey \textit{Pristine} \citep{PristineI} as a candidate extremely metal-poor star. It was selected as a candidate for follow-up spectroscopy following the procedure outlined in \citet{PristineIII}, based on its $CaHK$ filter magnitude from the \textit{Pristine} survey in combination with SDSS broad-band photometry (see Table \ref{phot}). Its photometric metallicity from \textit{Pristine} Ca H\&K narrow-band photometry and SDSS broad-band photometry \citep[see for a detailed description of the procedure][]{PristineI} was estimated to be [Fe/H]$ \sim -3.2$, but we note that, at metallicities well below [Fe/H]$ = -3$, even the very metallicity sensitive \textit{Pristine} photometry loses its discriminative power, and follow-up spectroscopy is therefore needed to determine the final metallicity of the star and, of course, to establish its overall abundance pattern.

\begin{table*}
\centering
\begin{tabular}{llrrrl}
\hline
 & unit & value & uncertainty  & extinction & source\\
 & & &  & applied & \\
\hline
\hline
Right Ascension & (h:m:s J2000) & 14:47:30.73 & --& -- & SDSS\\
Declination & (d:m:s J2000) & +09:47:03.70 & -- & -- & SDSS\\ 
\hline
$u_{0}$ & (mag) & 17.411 & 0.010 & 0.100 & SDSS\\
$g_{0}$ & (mag) & 16.512 &  0.004 & 0.078 & SDSS \\
$r_{0}$ & (mag) & 16.177 &  0.005 & 0.054 & SDSS \\
$i_{0}$ & (mag) & 16.035 & 0.005 & 0.040 & SDSS \\
$z_{0}$ & (mag) & 15.982 & 0.007 & 0.030 & SDSS\\
$CaHK_{0}$ & (mag) & 16.869 & 0.005 & 0.093 & \textit{Pristine}\\
$V_{0}$ & (mag) & 16.356 & 0.020 & 0.072 & APASS\\
$J_{0}$ & (mag) & 15.145 & 0.008 & 0.020 & UKIDSS to 2MASS\\
$H_{0}$ & (mag) & 14.779 & 0.011 & 0.010 & UKIDSS to 2MASS\\
$K_{0}$ & (mag) & 14.473 & 0.012 & 0.008 & UKIDSS to 2MASS\\
\hline
proper motion RA & (mas/yr) & $-7.76$ & 0.11 & -- & Gaia DR2 \\
proper motion DEC & (mas/yr) & $ -0.06$ & 0.12 & -- & Gaia DR2\\
parallax & (mas) & 0.119 & 0.094 & -- & Gaia DR2\\
\hline
distance & (kpc) & 6.9 & 0.3 & -- & isochrone fitting + Gaia DR2 \\
log(g) & (dex) & 3.5 & 0.5 & -- & isochrone fitting + Gaia DR2 \\
\teff\ & (K) & 5792 & 100 & -- & 3D corrected stellar models \\
\hline
\hline
\end{tabular}
\caption{Position, photometry from SDSS DR13 \citep{SDSS16} and \textit{Pristine}, astrometry from Gaia DR2 \citep{Brown18}, and adopted stellar parameters for \pris. \label{phot}}
\end{table*}

\subsection{Derivation of stellar parameters} \label{sec:stelpar}

\begin{figure}
\centering
\includegraphics[width=\linewidth]{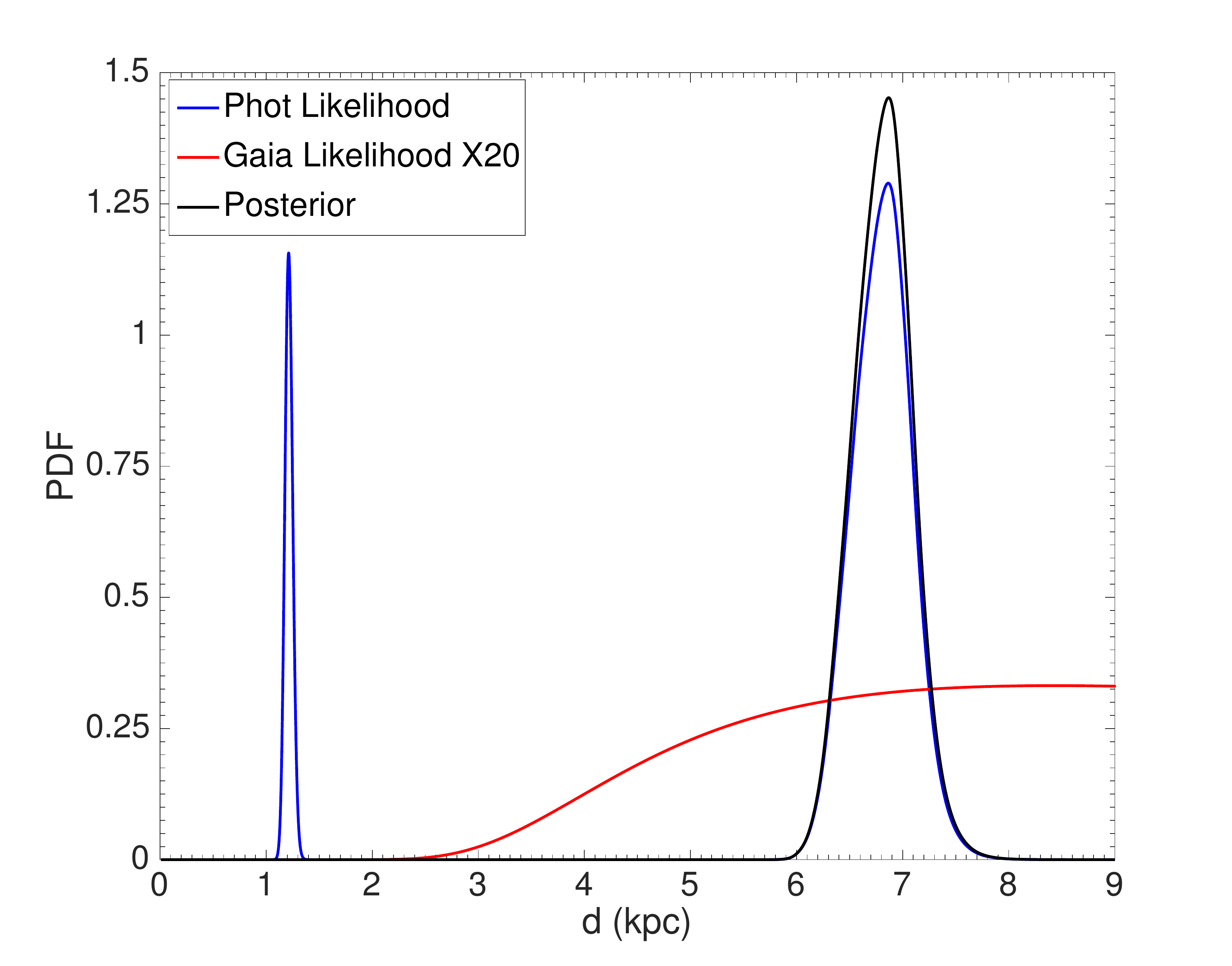}
\caption{Probability distribution function of the distance of \pris. Shown here are the likelihood based on the MESA isochrones alone (blue line), the likelihood based on the Gaia parallax (red line, multiplied by a factor 20 for visibility), and the combined posterior (black line). The sub-giant solution with a distance of $6.9 \pm 0.3$ kpc is favoured with a very high confidence level (99.7\%). \label{fig:par}}
\end{figure}

Table \ref{phot} summarises the photometry for Pristine\_221.8781+9.7844 that is available from SDSS DR13 \citep{SDSS16}. Comparison of the SDSS colours to the most metal-poor MESA isochrones \citep{Paxton11, Choi16, Dotter16} suggest that the star is either a sub-giant of log(g) $\sim$ 3.5 at a distance of $\sim$7 kpc, or on the main-sequence with log(g) $\sim$ 4.5 at a distance of $\sim$1.2 kpc. For the probability distribution function of the two solutions, shown by a blue line in Figure \ref{fig:par}, we have included the assumption that the star is old (age $>$ 11 Gyr), but the precise age is treated by a flat prior. Additionally, we have assumed that this extremely metal-poor star follows the halo density distribution described as a single power-law with $n=-3.4$ \citep[e.g.,][this is an average value, we have verified that the results are robust to a change in slope from $-2$ to $-4.5$]{Kafle14}. The resulting probability distribution favours the sub-giant solution for \pris\ by a factor 8, which is indicative, but can not be taken as a definitive answer. A careful fitting of the Balmer line series brought no clarity in this issue, as no combination of Teff and log(g) simultaneously provided a good fit to the full Balmer series. However, this degeneracy is broken by the parallax measurements from the ESA Gaia satellite released in Gaia DR2 \citep{Brown18} presenting a parallax of 0.119 $\pm$ 0.094 mas, thereby favouring the sub-giant solution with extremely high probability (99.7\%, see Figure \ref{fig:par}, details of the method to be published in Sestito et al., in prep.), and constraining the distance to \pris\ to 6.9 $\pm$ 0.3 kpc.

Using ($g_{0}-z_{0}$) = 0.530 and log(g) = 3.5, we derive \teff\ from an interpolation between the 3D-corrected ATLAS colours given in \citet{Bonifacio18b}, which gives our final adopted value of 5792 K. For comparison, we would derive \teff\ = 5805 K from the formula given in \citet{Ludwig08} for extremely metal-poor stars. The photometric calibration given by \citet{Casagrande10} yield a consistent result if we convert the SDSS magnitudes to ($V-I$) using the relation from \citet{Jordi06}. Unfortunately, the available photometry in the infrared bands from 2MASS \citep{Skrutskie06} has too large uncertainties to use for reliable effective temperatures. However, we instead use UKIDSS JHK magnitudes \citep{Lawrence12} transformed to the 2MASS system and de-reddened using the maps from \citet{Schlegel98}. Using the converted infrared magnitudes and the Johnson V magnitude from APASS \citep{Henden12}, we apply the infrared flux method \citep{Gonzalez09} to obtain \teff\ of 5877 $\pm$ 62 K. The method by \citet{Gonzalez09} is calibrated down to ultra-low metallicities ([Fe/H]$ = -4$) and the result is nicely consistent with our derived temperature from optical colours. Finally, we mention that the derived temperature value from Gaia DR2 photometry alone \citep{Andrae18} is also consistent although it has significantly larger uncertainties (\teff\ = 5862$^{+175}_{-169}$). 

The measured proper motion from Gaia DR2 is (pmRA, pmDEC) = ($-7.76 \pm 0.11$ mas/yr,$ -0.06 \pm 0.12$ mas/yr). Taken together with our measured radial velocity of $-$149 \kms\ from the spectra (see Section \ref{sec:hr}), this gives: ($U,V,W$)$ = $($-226.4^{+12.2}_{-15.8}, -171.6^{+14.1}_{-16.7}, 17.4^{+11.7}_{-8.4}$) \kms, with respect to the Galactic standard of rest. This 3D velocity is inconsistent with a star co-rotating in the Galactic disk, therefore we can conclude that it is a halo star. 

Table \ref{phot} summarises the adopted stellar parameters for \pris\ which will be used in the remainder of this work.

\section{Medium-resolution Spectroscopy}\label{sec:mr}

Initially, follow-up medium-resolution spectroscopy was obtained
with the Intermediate dispersion Spectrograph and Imaging System
(ISIS) \citep{Jorden90} spectrograph on the 4.2m William
Herschel Telescope (WHT) at the Observatorio del Roque de los
Muchachos on La Palma, Spain. We used the R600B and R600R
gratings, the GG495 filter in the red arm, and the default dichroic
(5300 \AA). The mean FWHM resolution with a one arcsecond
slit was R $\sim$2400 in the blue arm\footnote{The WHT observing setup was identical to that used in
\citet{Aguado16,Aguado17}.}. The observations were carried out over the
course of a five night observing run (15--19 July, 2017, Programme C31). Eight exposures
of 1800 s each were taken, although unfortunately with fairly high particle counts in the air 
and some of them at high airmass. A standard data reduction procedure including bias subtraction, flat-fielding,
and wavelength calibration using CuNe and CuAr lamps was performed
with the {\tt onespec} package in IRAF. The final signal-to-noise ratio of the
average reduced spectrum is S/N $\sim$180 at 4500 \AA.

\begin{figure*}
\centering
\includegraphics[width=\linewidth]{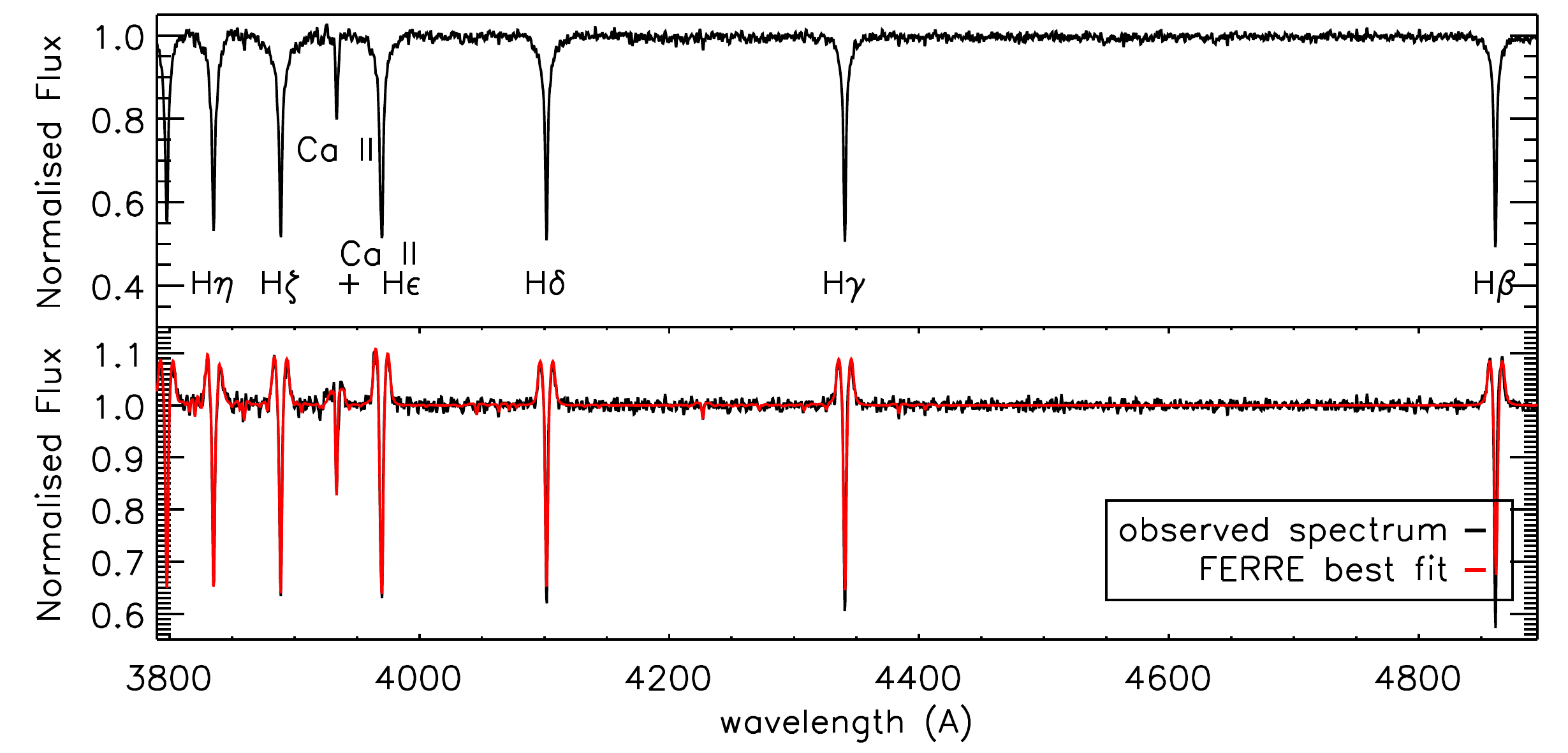}
\caption{Top panel: The medium-resolution spectrum for \pris\ as observed with WHT/ISIS. The main lines are labelled and one can clearly see that the medium-resolution spectrum is dominated by the H Balmer series. The weaker Ca II line is the Ca II K line, the Ca II H line is blended with H$\epsilon$. Bottom panel: The same spectrum, but now normalised by a running mean by the FERRE code and with the best-fit synthetic spectrum -- with the same normalisation -- overplotted in red. \label{fig:mr}}
\end{figure*}

Figure \ref{fig:mr} shows the spectrum, which is characterised by very few metal absorption 
lines and a very weak Ca II K line, which has an equivalent width of only $\sim$450 m\AA. The Ca II H line, to the immediate right, looks much stronger in 
comparison because it is blended with the Balmer H$\epsilon$ line. 

\subsection{Analysis}

To derive stellar parameters and chemical abundances, a grid of synthetic spectra was computed with the ASS$\epsilon$T
package \citep{Koesterke08} which uses the Barklem codes \citep{Barklem00a,Barklem00b} to describe the broadening of the Balmer lines. This grid is identical to that used and made publicly available by \citet{Aguado17}. The model atmospheres were computed with the same Kurucz codes and methods described by \citet{Meszaros12}. The
abundance of $\alpha$-elements was fixed to [$\alpha$/Fe] = +0.4, and the
limits of the grid were taken to be $-6 <$ [Fe/H] $< -2$, $-1 <$ [C/Fe] $< +5$,
4750K $< T_{\rm{eff}} <$ 7000K and 1.0 $<$ log g $<$ 5.0, with an assumed
microturbulence of 2 km~s$^{-1}$. We search for the best fit model using
FERRE\footnote{Available from github.com/callendeprieto/ferre} \citep{Allende06} by simultaneously deriving surface gravity, metallicity and carbon abundance. The observed and synthetic spectra were both normalized using a running-mean filter with a width of 30 pixels (about 10 \AA), see for the fit the lower panel of Figure \ref{fig:mr}. 

\subsection{Results}

The analysis with FERRE returns a best-fit value of [Fe/H] = $-$4.45 $\pm$ 0.21, an effective temperature of T$_{\rm{eff}}$ = 5871 $\pm$ 80 K, and a log(g) of 4.39 $\pm$ 0.5. We note that, in metal-poor stars, the surface gravity is the most difficult parameter to derive from medium-resolution spectroscopy. In this case FERRE has converged on the main-sequence solution for the evolutionary stage of the star, which is not supported by the Gaia DR2 parallax as shown in Figure \ref{fig:par}. The metallicity derivation is however not very sensitive to the correct log(g) value (see also Section \ref{sec:logg}). In addition, we note that adopting a different microturbulence value only marginally changes the derived parameters, well below the level of the given uncertainties \citep[see Section \ref{sec:logg} and more detailed tests in][]{Aguado18a}.

Due to the extreme weakness of most of the absorption lines, most information for the metallicity determination in this spectrum comes from the Ca II K line. Upon close inspection, though, this line looks slightly asymmetric, indicating that at this resolution the \ion{Ca}{II} K line is blended with an interstellar Ca absorption feature and that the actual metallicity is even lower. A high-resolution spectrum, as discussed in Section \ref{sec:hr}, is needed to resolve these two features. 

\begin{figure*}
\centering
\includegraphics[width=\linewidth]{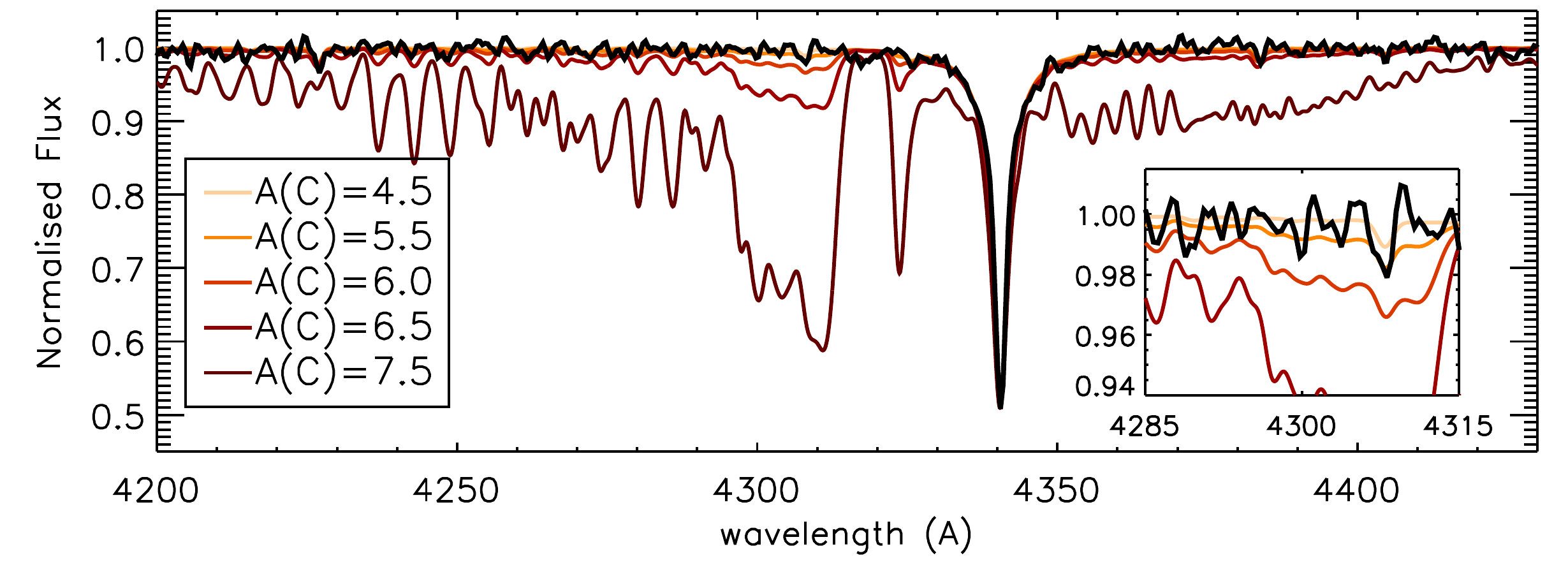}
\caption{ Synthetic spectra with \teff\ = 5792 K, log(g) = 3.5, [Fe/H] = -4.66 and different carbon abundances compared in the region of the carbon G-band to the low-resolution spectrum from of Pristine\_221.8781+9.7844 from the WHT. All synthetic spectra are smoothed to resolving power 2400 and the N and O abundances are changed in lockstep with the carbon abundance. For this purpose, the WHT spectrum has been continuum normalised locally (on a scale a little larger than the region shown here) by a linear function only to avoid any dipping of the continuum normalising function into the broad G-band features. The inset zooms in on the wavelength region most sensitive to the CH features. \label{fig:carbonlr}}
\end{figure*}

As illustrated in Figure \ref{fig:mr}, the absence of a carbon-band around 4300 \AA\ is a striking feature of the medium-resolution spectrum of \pris. This suggests that this star might be very carbon-poor in comparison to stars of similar [Fe/H]. In Figure \ref{fig:carbonlr}, we compare this high S/N medium-resolution spectrum with synthetic spectra of different carbon abundances in the region of the molecular G-band sensitive mostly to the CH molecule. The synthetic spectra are produced using MARCS (Model Atmospheres in Radiative and Convective Scheme) stellar atmospheres and the Turbospectrum code \citep{Alvarez98,Gustafsson08,Plez08} and adopt the stellar parameters from Table \ref{phot}. It is assumed that N and O change in lockstep with C, keeping [C/N] and [C/O] at the solar values, although we verified that changes in this assumption do not significantly influence the derived abundance. A subsequent analysis using instead ATLAS9 atmosphere models and the MOOG synthetic spectrum code to check for systematic effects yielded the same results. Extra care was taken in normalizing the observed spectrum by the continuum. To prevent any dipping of the continuum fitting into the broad carbon features, this spectrum was normalized locally (on a scale slightly larger than the wavelength range shown in Figure \ref{fig:carbonlr}) by a linear relation only. It is clear that the spectrum is not carbon-enhanced to the level of A(C) = 6.0 or above, and it is unlikely it is carbon-enriched to the value of A(C) $>$ 5.5, but it is difficult to constrain if the spectrum could be enhanced to a lower level. To quantify these results, we have run a set of MCMC experiments of the $\chi^2$ minimalisation by the FERRE code, using 10 Markov chains and 48000 Monte Carlo experiments \citep[similar to the method used in][]{Aguado18b}. The resulting distribution of outcomes peaks at A(C) = 5.15 (which corresponds to [C/Fe] = +1.10 as we adopt  the solar value A(C) = 8.50 by \citealt{Caffau10}, and [Fe/H] = $-$4.45 from FERRE). But, as is illustrated in Figure \ref{fig:carbonlr}, the results are more constraining at higher A(C) than at lower A(C) where all observable features vanish, and we thus regard the results as informative mostly on the upper limit of detectable carbon. In 68\% of the runs, the resulting value is less than A(C) = 5.2, in 95\% of the runs it is less than A(C) = 5.6. We adopt the latter as a robust upper limit.

\section{High-resolution Spectroscopy}\label{sec:hr}
After an analysis of the medium-resolution spectrum, we were allocated four hours of Director's Discretionary time
on the ESO/VLT (Programme {\tt 299.D-5042}) to obtain high-resolution spectroscopy using the UVES spectrograph \citep{Dekker00}.
The observations were split in four observing blocks, each of one hour and corresponding to 3005\,s of total integration time. We chose to use the standard setting DIC1 390+580, that covers the wavelength intervals 3300 \AA\ -- 4500 \AA\ in the blue arm, and 4790 \AA\ -- 5760 \AA\ and 5840 \AA\ -- 6800 \AA\ in the red arm, which was combined with a slit width of 1\farcs{2} with $1\times 1$ binning on the CCD. The observing blocks were executed in service mode,  when the star was close to the meridian, at the beginning of the nights of 14--17 August, 2017.

The spectra were reduced using the ESO Common Pipeline Library, UVES pipeline version 5.8.2. The reductions included 
bias subtraction, background subtraction optimal extraction, flat-fielding of the extracted spectra, wavelength calibration
based on the spectrum of a Th-Ar lamp, resampling at a constant wavelength step and optimal merging of the echelle orders. 
The reduced spectra were then corrected for the barycentric velocity.  

On the night of August 16, poor seeing conditions resulted in a slit-width-dominated resolution of the spectrum for that night, which is also reflected 
in a lower signal-to-noise ratio. To combine all spectra, they are brought to the rest wavelength, smoothed to a common 
resolution of 30\,000 (taking into account the slightly different resolution in the spectrum from August 16) and combined by summation. 
The radial heliocentric velocity for the star is measured to be --149.0 $\pm$ 0.5 \kms; no significant velocity variations are measured 
on different nights. The approximate signal-to-noise ratios per pixel of the final combined spectrum are 45 at 4000 \AA, 50 at 4300 \AA, 85 at 5200 \AA, and 100 at 6700 \AA.

\subsection{Analysis} \label{sec:ana}
For the high-resolution analysis, we take full advantage of the expertise within the Pristine team and analyse the spectrum using four different techniques. This approach, not uncommon among modern surveys \citep[e.g.,][]{BailerJones13, Smiljanic14}, allows to quantify and fold in different sources of uncertainties, including systematic uncertainties --- due to measuring technique, continuum placement, model atmospheres, or adopted synthetic spectrum code ---  to give a robust measurement of the precision of the abundances. The four methods are briefly described below. They vary widely in their approach. For instance, two methods are equivalent width based and two instead rely on spectral fitting. Different model atmospheres and synthetic spectral codes are used. To make sure we operate on a common scale however, all four methods use the line list compiled by C. Sneden for the synthesis of spectra using the code MOOG \citep[2016, private communications, derived from][]{Kurucz95}; updated \ion{Fe}{I} atomic line data are additionally implemented from \citet[][]{OBrien91},\citet{Wood13}, and \citet{DenHartog14}. The full line list is presented in Appendix A. In all analyses, we use the solar abundances from \citet{Lodders09} with C and Fe solar abundances taken from \citet{Caffau11b}. All methods use 1D LTE approaches and the same  stellar parameters, we have adopted the sub-giant branch log(g) of 3.5 and the corresponding 3D model temperature of 5792 K as given in Table \ref{phot}. 

\subsubsection{Method 1}
Using the same stellar grid as for the medium-resolution spectral analysis, we normalise both the grid and the UVES spectra using a running mean filter of 500 pixels-per-window. In these spectra, 184 windows, each of width 2 \AA, are identified around the strongest metallic absorption lines. We subsequently use FERRE to derive the individual abundance of every single line. Based on the $\chi^{2}$ value of every fit and following a visual inspection, 53 lines were deemed to be reliable. The lines are indicated in Table \ref{tab:linelist} in Appendix A. 

\subsubsection{Method 2}
We additionally analyse the spectra using the \mygi\ code \citep{mygi} following Paper II of this series \citep{PristineII}. \mygi\ directly fits the synthetic profile of each chosen feature against the observed one using a pre-computed grid. Here we use a  grid  of 126 ATLAS\,12 models \citep{Kurucz05,Castelli05} with temperatures between 5600\,K and 6100\,K and a step of 100\,K, surface gravities log g = 3.5, 4.0, 4.5 and metallicities between --5.0 and --3.25 at steps of 0.25\, dex. For all the models we assume solar scaled abundances with $\alpha$-elements enhanced by 0.4\,dex, a mixing length parameter 1.25, and microturbulent velocity of 1.0\,\kms. From this grid of models we use SYNTHE \citep{Kurucz05}, in its Linux version \citep{Sbordone04,Sbordone05}, to compute a grid of 378 synthetic spectra. In the analysis for this work the microturbulent velocity is fixed at 1.5 km/s. 

\subsubsection{Method 3}
This method uses a classical equivalent width approach in which the equivalent widths are measured with DAOSPEC \citep{Stetson08} through the wrapper 4DAO \citep{Mucciarelli13a}. Uncertainties on the equivalent width measurements are estimated by DAOSPEC as the standard deviation of the local flux residuals. All the lines with equivalent width uncertainties larger than 15\% are excluded from the analysis. Lines fainter than 10 m\AA\  are discarded as well. Chemical abundances are estimated from the measured equivalent widths by using the package GALA \citep{Mucciarelli13b}. We run GALA keeping \teff, log(g) and microturbulence of the model fixed, allowing its metallicity to vary iteratively in order to match the Fe abundance measured from equivalent widths. All model atmospheres are computed with the ATLAS9 code \citep{Castelli04}; the results presented in Section \ref{sec:abu} assume a microturbulence value of 1.5 km~s$^{-1}$. 

\subsubsection{Method 4}
Our final method uses equivalent width measurements from integrating the area under the continuum using IRAF. Lines with measurement uncertainties over 15\% were removed. The spectrum synthesis code MOOG was used for the abundance analysis \citep[2014 version][]{Sneden73}.  Model atmospheres were adopted from \citet{Plez02}, which include LTE, plane-parallel models, with [Fe/H] $\ge -4.0$, and [$\alpha$/Fe] = +0.4. Additionally, [C/Fe] = 0 was adopted. Microturbulence was determined per model from the MOOG analysis using the many available \ion{Fe}{I} lines, by requiring no relationship between Fe abundance and line strength.  The uncertainty in the microturbulence values is estimated as $\pm$0.1 km/s.

\section{The abundance pattern of \pris\ } \label{sec:abu}

\begin{table*}
\centering
\begin{tabular}{lllllllllllll}
\hline
Ion  & Method1 & $\sigma$1 & N1 & Method2 & $\sigma$2 & N2 & Method3 & $\sigma$3 & N3& Method4 & $\sigma$4 & N4 \\
\hline
\hline
\multicolumn{13}{c}{For stellar parameters \teff\ = 5792 K and log(g) = 3.5}\\
\hline
\ion{Na}{I} &  1.86 & 0.07 & 2 & -- & -- & -- & 2.24 & 0.03  &  2  & 2.20 & 0.04 & 2\\
\ion{Mg}{I}  & 3.30 & 0.15 & 6 & 3.39 & 0.12 & 4 &  3.37 & 0.07  &  3 & 3.38 & 0.09 & 6\\
\ion{Si}{I}  & 3.09 & -- & 1 & 3.26 & -- & 1 & 3.23 & 0.02  &  1  & 3.23 & -- & 1\\
\ion{Ca}{I}  & 1.68 & -- & 1 & 1.96 & -- & 1 &  1.80 & 0.02 &   1 & 1.91 & -- & 1\\
\ion{Ca}{II} & 2.19 & 0.29 & 3 & 2.26 & -- & 3 & -- & -- & -- & -- & -- & --\\
\ion{Ti}{II}  & 0.96 & 0.44 & 2 & 0.62 & -- & 1 &  0.65 & 0.003  &  2  & 0.78 & 0.01 & 2\\
\ion{Fe}{I} & 2.79 & 0.19 & 35 & 2.88 & 0.18 & 29 &  2.84 & 0.15  & 35 & 3.01 & 0.17 & 36\\
\hline
$v_{\rm{turb}}$ (\kms) & 2.0 & & & 1.5 & & & 1.5 & & & 1.3 & & \\
\hline
\end{tabular}
\caption{Chemical abundances derived by the several methods. We also list the microturbulence derived (in the case method 4), or assumed (for methods 1, 2 and 3), by each of the methods and for each set of stellar parameters. \label{tab:abu}}
\end{table*}

\begin{figure}
\centering
\includegraphics[width=\linewidth]{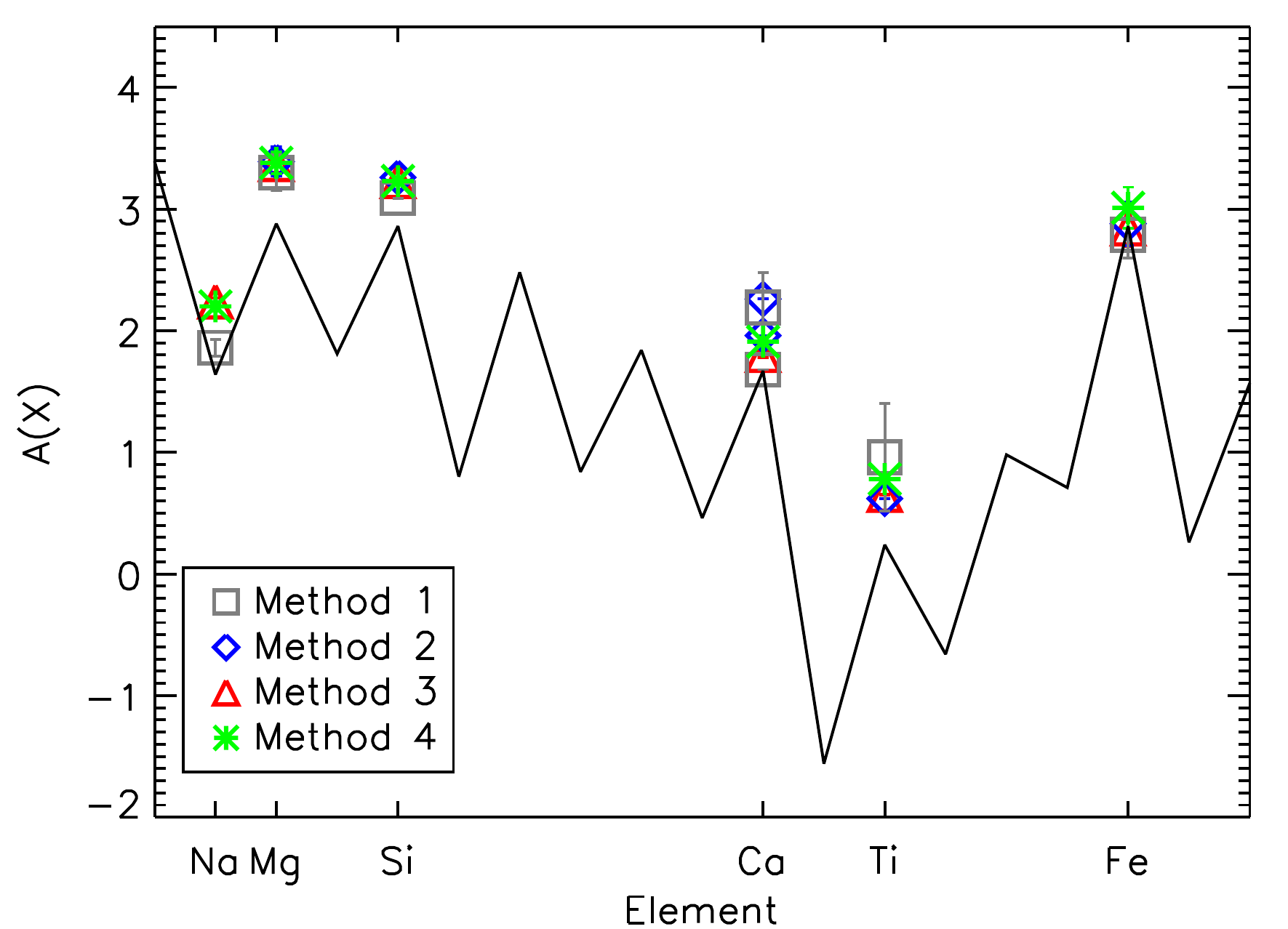}
\caption{Abundances derived by all four methods assuming  \teff\ = 5792 K and log(g) = 3.5, overplotted by a shifted solar abundance scale with [Fe/H] = $-$4.66 following the solar abundances from \citet{Lodders09} with the Fe solar abundance taken from \citet{Caffau11b} \label{fig:abu}}
\end{figure}

\subsection{Fe, Na and the $\alpha$-elements}\label{sec:logg}

We use all four methods described in Section \ref{sec:ana} to measure the abundances of individual elements in the high-resolution UVES spectrum as summarised in Table \ref{tab:abu} and shown in Figure \ref{fig:abu}. The overall 1D LTE abundance pattern in Fe, Na and $\alpha$-elemental abundances is relatively well described by a scaled-solar abundance pattern for a star with [Fe/H] = $-$ 4.66, as shown by the comparison of the symbols with the solid line in Figure \ref{fig:abu}. We do observe a small enhancement in the $\alpha$-elements (most notably Mg) relative to this pattern, this is not uncommon among metal-poor stars. All $\alpha$-elements are elevated to similar [X/Fe] levels. The agreement between the methods is very good, and any remaining disagreements might be related to details in continuum placement and the use of different atmosphere models and synthesis codes. However, the scatter is reassuringly small for all of the measured element abundances.  

\begin{table*}
\centering
\begin{tabular}{lllllll}
\hline
El & Teff+ 200 K & Teff - 200 K & logg + 0.5 dex & logg $-$0.5 dex & vt +0.5
\kms  & vt $-$ 0.5 \kms   \\
\hline
\hline
Na  &  $+$0.150 &  $-$0.157  & $-$0.004  & $+$0.010 &$-$0.015  &  $+$0.016 \\
Mg   & $+$0.137 &  $-$0.143 &  $-$0.002  & $+$0.006  &  $-$0.020 & $+$0.021 \\  
Si   & $+$0.166  & $-$0.169  &  $+$0.023 &  $-$0.007  &  $-$0.017 &   $+$0.017   \\
Ca  & $+$0.182  & $-$0.189 &  $-$0.004  & $+$0.012  &  $-$0.033 & $+$0.041   \\ 
Ti  &  $+$0.107 &  $-$0.111  &  $+$0.168 &  $-$0.162  &  $-$0.035 &   $+$0.048    \\
Fe  &  $+$0.207 &  $-$0.204 &   $+$0.002 & $+$0.010  &  $-$0.046 &  $+$0.097   \\
\hline
\end{tabular}
\caption{Changes of the derived abundances depending on stellar parameters}\label{tab:uncstelpar}
\end{table*}

\begin{table}
\centering
\begin{tabular}{llrrrr} % four columns, alignment for each
\hline
Ion & A(X)$_\odot$ & A(X) & [X/H] & $\sigma$ & [X/Fe] \\
\hline
\hline
\ion{Li}{i} & 1.10 & 1.70 & 0.60 & 0.20 & 5.26 \\
\ion{C}{i} & 8.50 & $<$5.60 & $< -$2.90 &  & $<$1.76 \\ 
\ion{Na}{i} & 6.30 & 2.20 & $-$4.10 & 0.10 & 0.56  \\ 
\ion{Mg}{i} & 7.54 & 3.37 & $-$4.17 & 0.08 & 0.49 \\
\ion{Al}{i} & 6.47 & $<$1.60 & $< -$ 4.87 &  & $<-$0.21\\
\ion{Si}{i}  & 7.52 & 3.23 & $-$4.29 & 0.09 & 0.37  \\
\ion{Ca}{i}  & 6.33 & 1.86 & $-$4.47 & 0.12 & 0.19 \\
\ion{Ca}{ii} & 6.33 & 2.22 & $-$4.11 & 0.14 & 0.55 \\
\ion{Ti}{ii} & 4.90 & 0.70 & $-$4.20 & 0.20 & 0.46 \\
\ion{Fe}{i}  & 7.52 & 2.86 & $-$4.66 & 0.13 & 0.00 \\
\ion{Sr}{ii} & 2.92 & $<-$1.50 & $<-4.42$ &  & $<$0.24\\
\hline
\end{tabular}
\caption{Derived final 1D LTE abundances for Pristine\_221.8781+9.7844. We adopt the stellar parameters \teff\ = 5792 K, log(g) = 3.5, and $v_{\rm{turb}}$ = 1.5 \kms. \label{abund}}
\end{table}

In Table \ref{tab:uncstelpar}, the changes in abundances are given as a function of the stellar parameters \teff, log(g), and microturbulent velocity. These changes were calculated with Method 3, but are in good agreement with similar analyses with any of the other methods. Changes in the adopted microturbulent velocity result in very small abundance changes for all elements, but can explain some of the discrepancies between the methods, such as for instance in the [Fe/H] values derived by method 4 and 1. On the other hand, the [Fe/H] value of the star is not very sensitive to surface gravity, as it is measured through Fe atoms in the neutral state. This is the case for most abundances with the most notable exception being \ion{Ti}{II}, which does show a much stronger abundance change with a change in surface gravity. Most elements show a significant change with a 200 K change in temperature.

The individual abundances from each method were combined to yield the final 1D LTE abundances in Table \ref{abund}. To obtain these combined values we assume that the measurements from each method are drawn from an unknown normal distribution $\mathcal{N}$($\mu,\sigma$). The other assumption we make is that each abundance measurement is accompanied by a Gaussian measurement uncertainty, $e_{i}$. The uncertainties for each measurement given in Table 3 are quite inhomogeneous. For example, in methods 1, 2, and 4 they only reflect the dispersion of the measurements, whereas in method 3 the uncertainty on the equivalent width measurement is additionally taken into account. Instead of adopting these inhomogeneous and incomplete uncertainties, we treat each measurement error as an unknown model parameter to be marginalized over. We use uniform priors in the [0,5] range for $\mu$, and scale-invariant priors for both $\sigma$ and $e_{i}$, i.e., $p(\theta) \propto \theta^{-1}$ for $-5 < ln\,\theta < 1$. We sample the posterior probability distribution function using an MCMC algorithm with $10^5$ steps, thereby discarding the first $10^4$ burn-in steps when exploring the posterior distribution (for a general description of this method see \citealt{Ivezic14}). The final abundance that we compute this way is more robust than a straightforward (weighted) mean of the measurement values. As this only reflects the measurements at a fixed set of stellar parameters (\teff, log(g), and microturbulence), we additionally add in quadrature to the derived $\sigma_{\rm{true}}$ the mean uncertainty corresponding to a change in stellar parameters of 100 K, 0.5 dex in log(g) and 0.5 in microturbulent velocity as derived from the values in Table \ref{tab:uncstelpar} to represent the typical uncertainties in these parameters. 

\subsection{Carbon abundance} \label{sec:c}

\begin{figure}
\centering
\includegraphics[width=\linewidth]{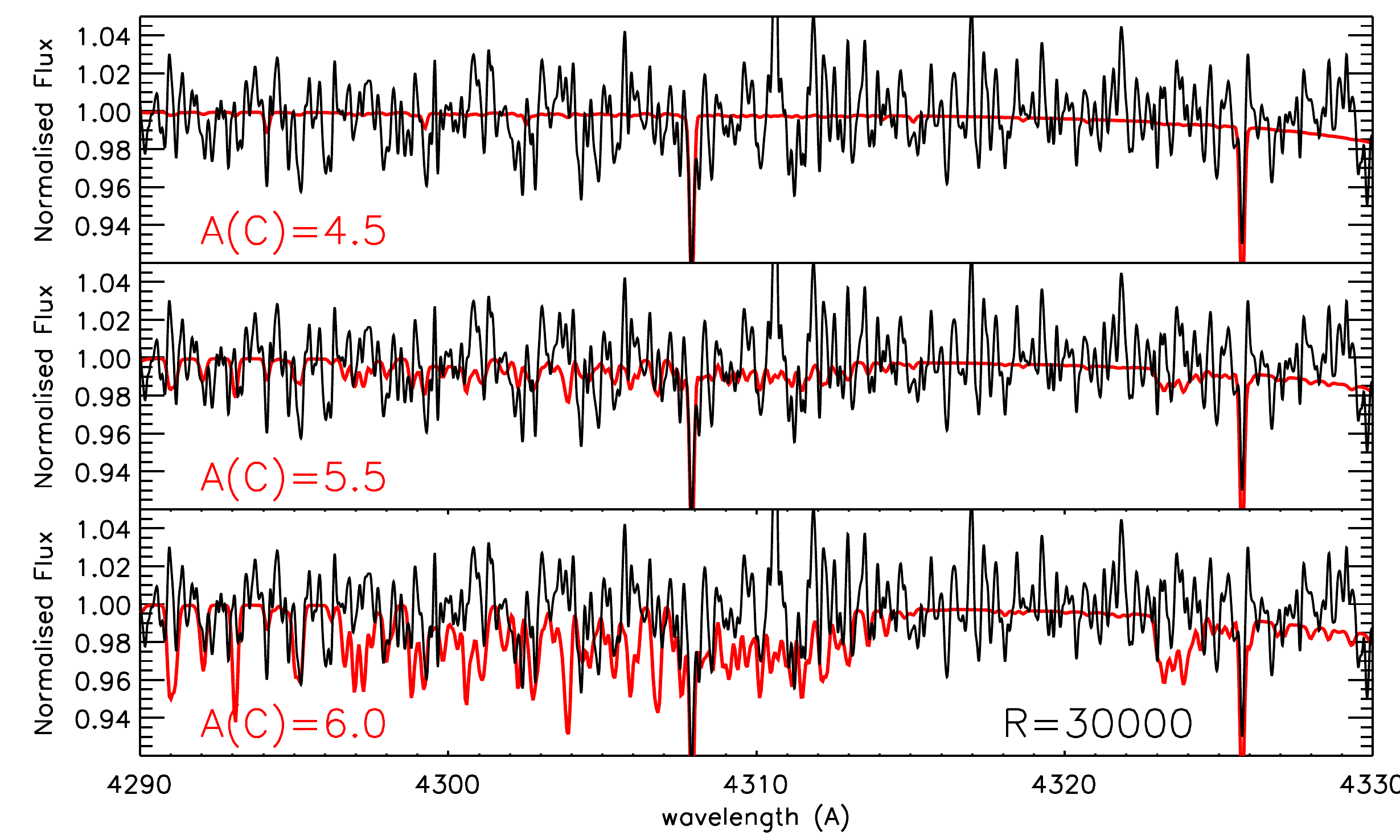}
\caption{Synthetic spectra with \teff\ = 5792 K, log(g) = 3.5, [M/H] = $-4.66$ (we assume that the other elements follow the solar pattern to first order and set [M/H] to [Fe/H]) and different carbon abundances (from top to bottom panel A(C) = 4.5, 5.5, or 6.5, corresponding to [C/Fe] = +0.66, +1.66, and +2.66) zoomed in relatively to Figure \ref{fig:carbonlr} on the most striking carbon G-band features, compared to the high-resolution spectrum from of Pristine\_221.8781+9.7844 from UVES. At the reddest wavelengths illustrated here, the onset of the broad H$\gamma$ line can be seen. All synthetic spectra are smoothed to resolving power 30,000 and the N and O abundances are changed in lockstep with the carbon abundance. Normalisation of the UVES spectrum has been compared to the WHT spectrum shown in Figure \ref{fig:carbonlr} to avoid any straightening of larger scale features.\label{fig:carbonhr}}
\end{figure}

In Figure \ref{fig:carbonhr}, we use the high-resolution UVES spectrum to look into the carbon features in more detail to investigate if we can put a more tight constraint on A(C) as was obtained from the medium-resolution spectrum in Section \ref{sec:mr}. Because UVES is an echelle spectrograph, the normalisation of the spectrum is more complex, but care has been taken that it matches the low-resolution spectrum when the spectrum is convolved to the same resolution, i.e. we carefully checked  that the normalisation does not follow any of the broader features, which would possibly erase the signature of the carbon band. Again, we see no evidence in this spectrum for specific carbon features. However, the signal-to-noise ratio is not sufficient to detect carbon enhancement below the A(C) = 5.6 level already set by the higher signal-to-noise medium-resolution spectrum. Guided by the synthetic spectra, the difference between A(C) = 5.5 and A(C) = 4.5 never exceeds 2\% of the flux, thus requiring a much higher signal-to-noise level than available. We adopt A(C)$ = 5.6$ as our final 1D LTE upper limit. This corresponds to [C/Fe] $= 1.76$, when taken together with the high-resolution measurement of iron, [Fe/H]$ = -4.66$.

\subsection{Lithium}\label{sec:li}

At metallicities higher than [Fe/H]$ = -3.0$, unevolved stars (i.e., in the main-sequence, turn-off, or sub-giant phases of stellar evolution)
display a constant lithium abundance: the familiar Spite plateau \citep{Spite82a,Spite82b}. At lower metallicities, though, there is a tendency to 
find lower lithium abundances. This is the so-called ``meltdown of the Spite plateau'' (\citealt{sbordone10}, but see also \citealt{bonifacio07,aoki09}). 

\begin{figure}
	\includegraphics[width=\linewidth]{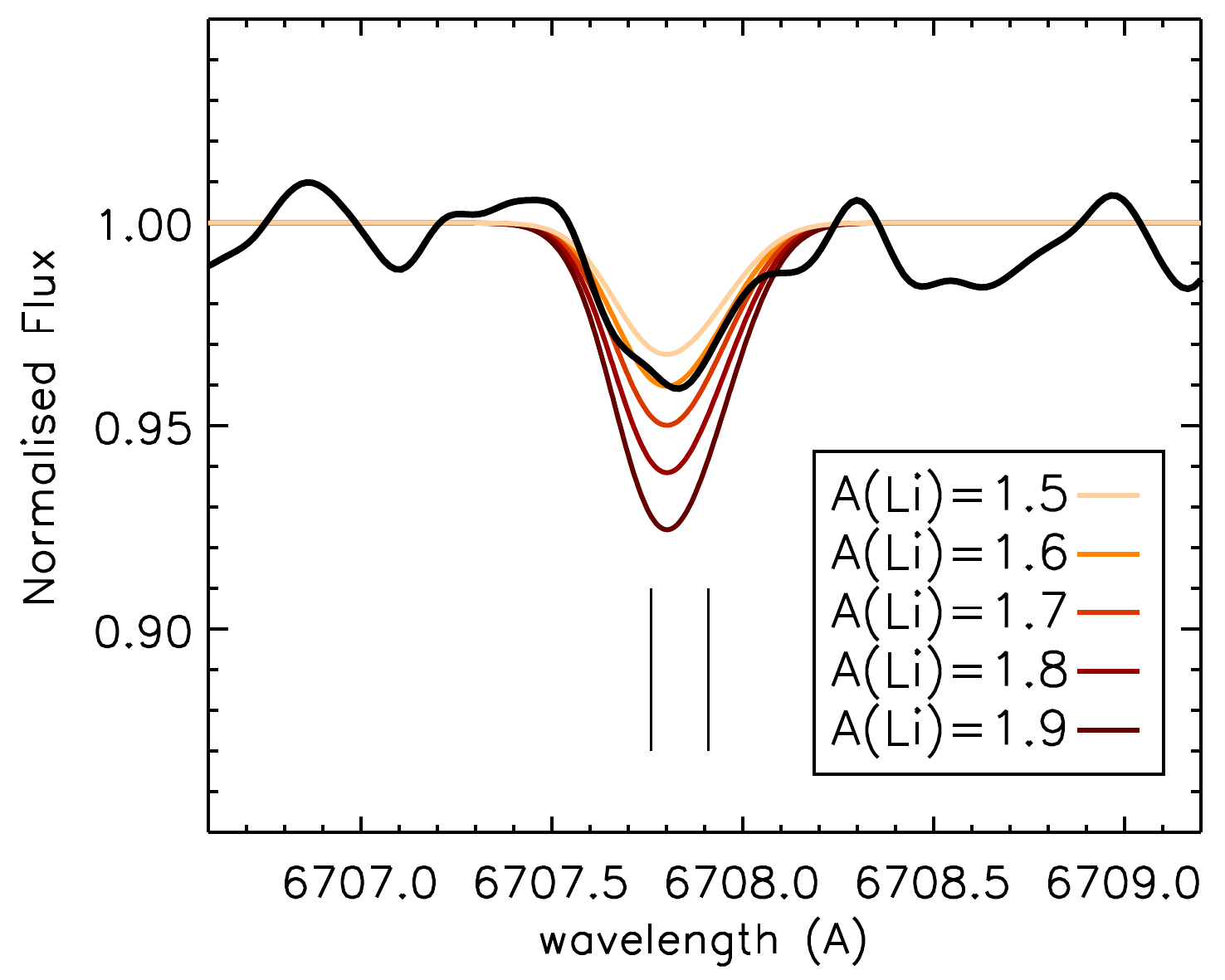}
    \caption{The \ion{Li}{i} resonance doublet in Pristine\_221.8781+9.7844 (black line) overplotted by synthetic spectra created from MARCS models and the Turbospectrum code as coloured lines. In the synthetic spectra the stellar parameters from Table \ref{phot} are used together with [Fe/H]$ = -4.66$. The linelist adopted for the small wavelength region shown in this figure is taken from \citet{Guiglion16}. The two vertical lines indicate the main components of the doublet that are not resolved at this resolution. The best fit 1D LTE lithium abundance is A(Li) = 1.6, this corresponds to A(Li) = 1.7 in 3D NLTE \citep[see][]{sbordone10}. \label{lithium}}
\end{figure}

\begin{figure}
	\includegraphics[width=\linewidth]{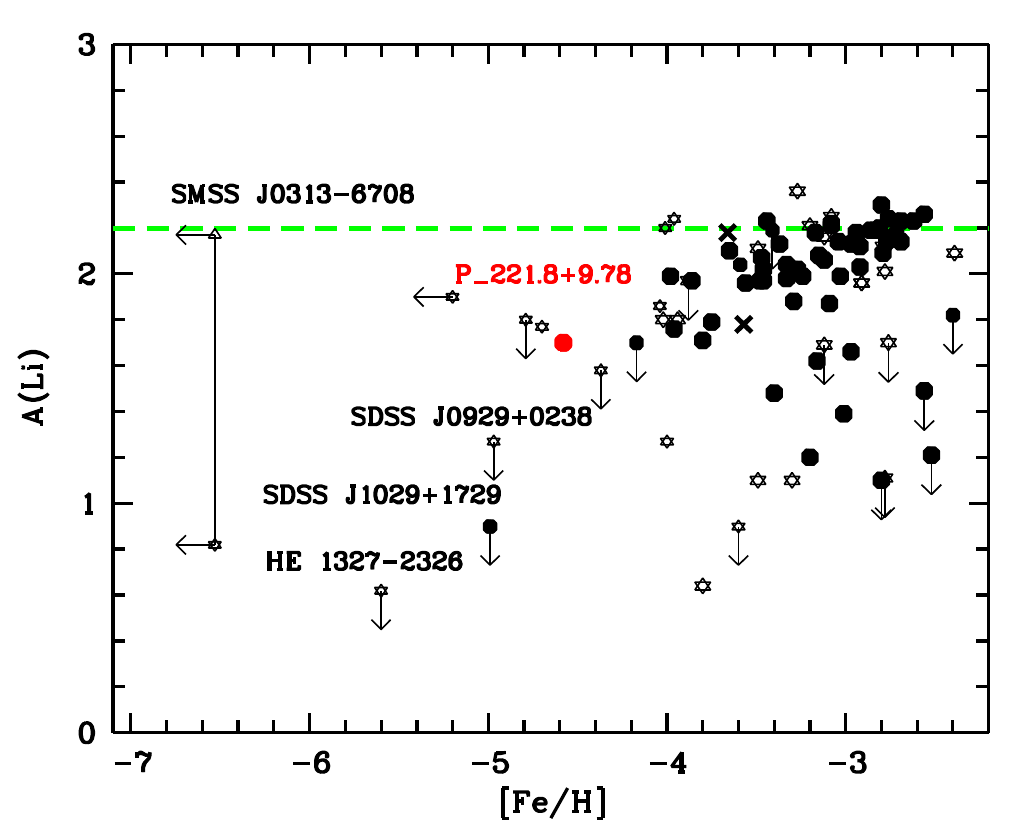}
    \caption{Lithium abundances as a function of [Fe/H] for unevolved stars. 
The filled black symbols are to carbon-normal stars and the open star-symbols
are low carbon band CEMP stars, as defined in \citet{Bonifacio18a}.
The literature values of lithium and carbon have been taken from: 
\citep{norris97,Lucatello03,
Ivans05,frebel07,Frebel08,thompson08,aoki08,
sbordone10,Caffau12,
Carollo12,masseron12,aoki13,
Ito13,Carollo13,Spite13,Roederer14,Aoki15,Bonifacio12,Li15,Hansen14,Caffau16,
Placco16,Matsuno17,Bonifacio18a}.
The two components of the binary system CS 22876-32 \citep{Gonzalez08} are shown as black crosses.
All the stars in this plot are unevolved, except for SMSS\,J031300.36-670839.3, which is a giant for which we take A(Li) and the upper limit on Fe
from \citet{Nordlander17}. The triangle, connected by a line to the measured
point of  SMSS\,J031300.36-670839.3 corresponds to the value corrected for 
dilution as computed by \citet{Mucciarelli12} for giant stars below the RGB bump.
The green dashed line is the level of the {\em} Spite plateau
as determined by \citet{sbordone10}. \label{life}}
\end{figure}

The \ion{Li}{i} resonance doublet at 6707\AA\ for \pris\ is shown in Figure \ref{lithium}.
We measure an equivalent width of 16.2 $\pm$ 0.7 m\AA\ which corresponds to A(Li) = 1.7 for \pris\ using the 3D-NLTE fitting formula of \citet{sbordone10}. The uncertainty derived from the equivalent width measurement uncertainty of 0.7 m\AA\ is very small ($\sim$ 0.02 dex), but we note that additionally this measurement is quite sensitive to continuum placement. By shifting the continuum, we retrieve a lower limit of 11 m\AA\ which would correspondingly lower A(Li) by 0.2 dex.
An adopted A(Li) = 1.7 would already place the star well below the Spite plateau, as shown in 
Figure \ref{life}.  Pristine\_221.8781+9.7844 is then the third unevolved star with  [Fe/H]$<-4.0$ 
and a measured lithium abundance. The other two unevolved stars with a Li measurement
are HE\,0233-0343 \citep{Hansen14}
and SDSS\,J1035+0641 \citep{Bonifacio18a}. These two stars are clearly carbon enhanced. The lithium abundance in these three stars is
similar. All other unevolved  stars  with  [Fe/H]$<-4.0$  have only upper
limits to their Li abundance.

For the giant star SMSS\,J031300.36-670839.3 with [Fe/H]$< -7.1$, \citet{Nordlander17}
measure A(Li)$ = 0.82\pm 0.08$. If we assume that the star is less
luminous than the RGB bump we can estimate the dilution using standard
models, as in \citet{Mucciarelli12} which would imply a Li
abundance (taking diffusion into account) at the turn-off for this star of A(Li) = 2.17.
This correction would place SMSS\,J031300.36-670839.3 squarely on the Spite plateau.

\citet{Bonifacio18a} noted that, among stars with [Fe/H]$<-3.5$ and \teff\ less than 6000\,K,
Li is always depleted, suggesting that the edge for Li depletion
becomes hotter for the more metal-poor stars. Pristine\_221.8781+9.7844 follows
this general behaviour.

\subsection{Other abundance constraints}
For several other elements where no detection could be established, we derived an upper limit. In the UVES spectrum, some Al lines are present but are very weak, and we derived from them an upper limit of A(Al) $\leqslant$ 1.6. The \ion{Sr}{II} line at 4077.7 $\AA$\ can be seen, but looks distinctly non-Gaussian in shape, which is why we also treat this line as an upper limit of A(Sr) $\leqslant -1.5$.  

\subsection{Non-LTE and 3D effects}\label{sec:nonlte}

With the exception of the Lithium abundance, all abundances derived in Section \ref{sec:li} do not take any non-LTE or 3D corrections into account. However, we note that SDSS J102915+172927, which has very similar stellar parameters as \pris\ (see Section \ref{sec:compare}) has been analysed using 3D and non-LTE for many elements, as described in \citet{Caffau12}. This, however, does not predict the combined 3D non-LTE correction strength as the two effects influence each other and a full 3D non-LTE correction cannot be achieved by adding a separate 3D correction and additionally a non-LTE correction. 

A complete 3D, non-LTE correction to the abundances is also beyond the scope of this paper but, because of these striking similarities between the two stars, we refer the reader to the 3D and non-LTE corrections derived for SDSS J102915+172927 as indicative of the magnitude of the corrections expected for the elements presented here. For SDSS J102915+172927, the 1D non-LTE correction for [Fe/H] is +0.13, while the 3D LTE correction is $-$0.27. Most significantly, the 3D LTE study impacts the [C/H] abundance ratio (see also Section \ref{sec:c}), which is corrected by $-$0.7 dex. No non-LTE calculations are available for this abundance derived from molecular CH. The 1D non-LTE calculations that are available for other elements in SDSS J102915+172927 show the largest correction for Si ($-0.34$ dex). The non-LTE corrections for neutral and once ionised Ca tend to go in opposite directions (each by $\sim$0.25 dex) and will make the discrepancy between the two abundances smaller in \pris. We further note that the corrections for 1D non-LTE typically increase with decreasing log(g) such that \pris\ log(g) = 3.5 needs a slightly larger correction than a log(g) = 4.0 star like  SDSS J102915+172927 \citep[see Table 4 of][]{Caffau12}, however, the differences in the corrections are not large enough to cover the difference in the [$\alpha$/Fe] pattern we observe for the two stars.  Overall, we can conclude that 3D and non-LTE corrections will improve our detailed picture of the abundances in \pris\, but that the global properties of the star will not be significantly affected.

\section{Comparison to the most metal-poor stars known}\label{sec:compare}

\begin{figure}
\centering
\includegraphics[width=\linewidth]{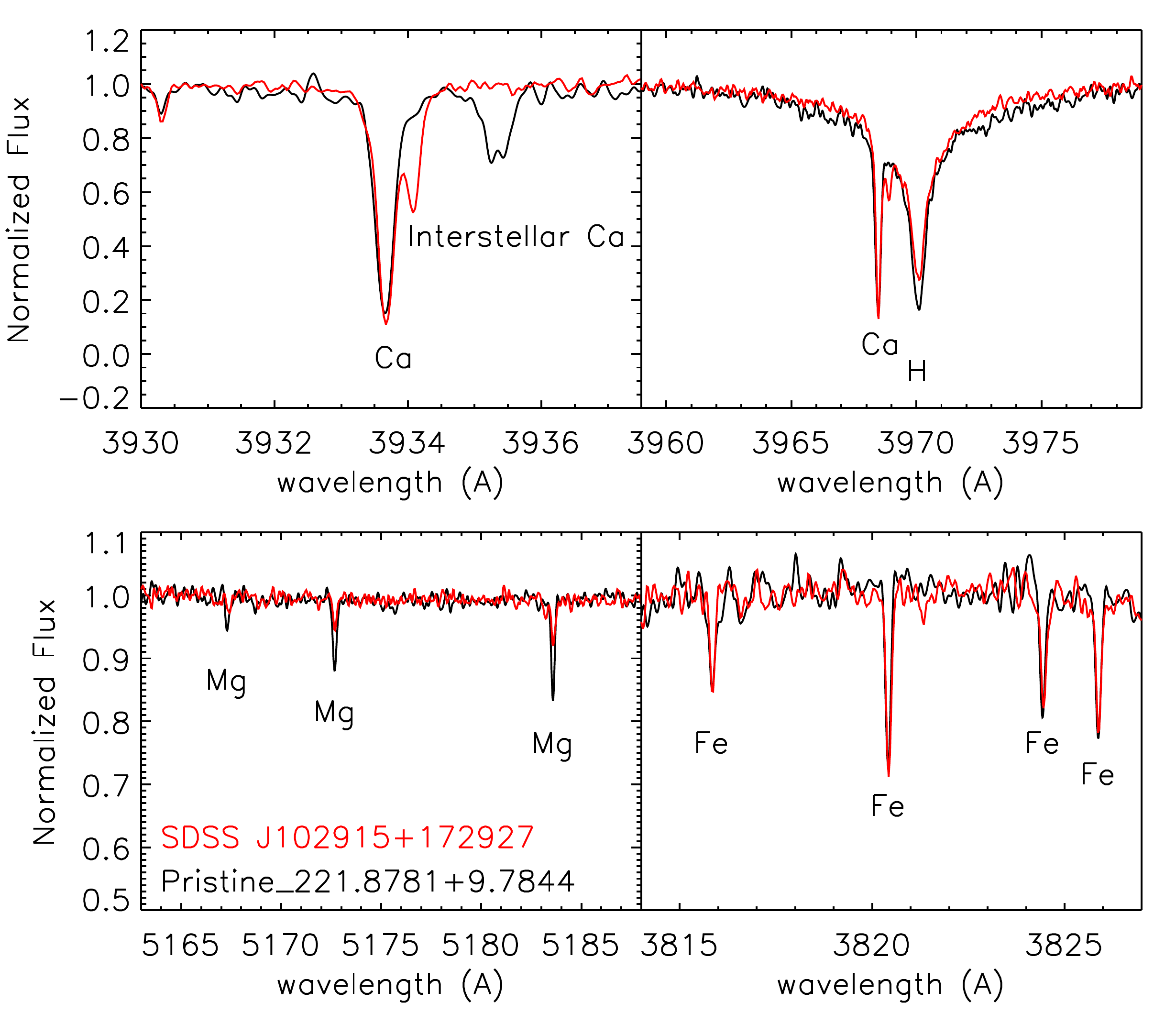}
\caption{Comparison of Pristine\_221.8781+9.7844 with the spectrum of SDSS J102915+172927. The spectrum of SDSS J102915+172927 is also taken with the same UVES setup and can thus be directly compared. From top left to bottom right the panels are centred on the Ca II K line, the Ca II H line, the Mg triplet, and a region with many Fe lines respectively. In the top left panel we see both the Ca II K absorption feature from the star itself as well as an interstellar component. Because the stars have different radial velocities and are shifted here to their stellar spectral rest wavelength, the interstellar components appear at different wavelengths in this figure. A great advantage of the high-resolution spectrum is that it resolves these two features, so that interstellar Ca absorption is not confused with stellar Ca.\label{fig:compSDSS}}
\end{figure}

\begin{figure}
\centering
\includegraphics[width=\linewidth]{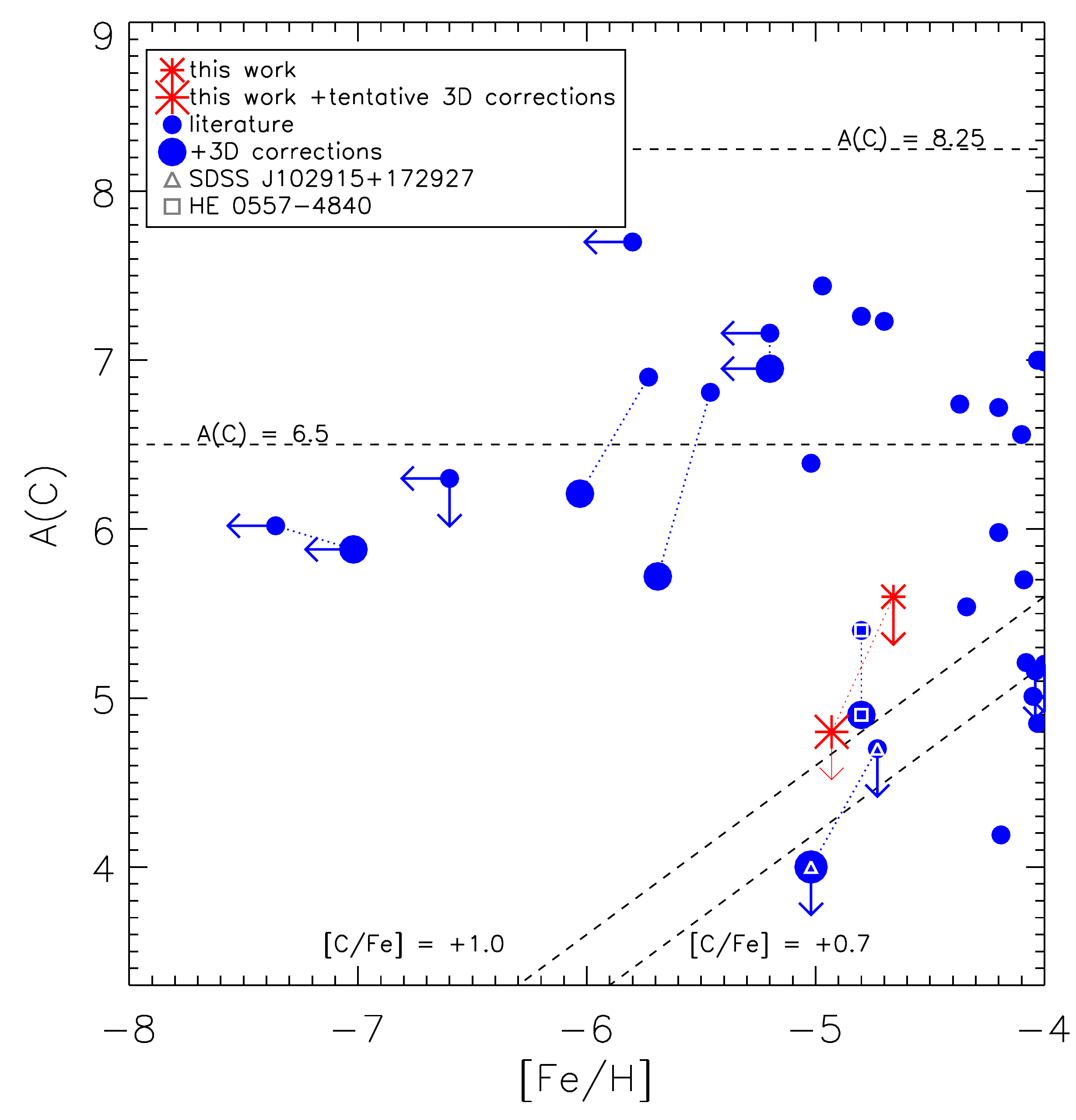}
\caption{The results of our abundance analysis for \pris\  from 1D LTE analysis (large red asterisk) and through face-value applications of the 3D corrections on carbon and iron from \citet{Caffau12} (larger red asterisk) compared to a literature sample as collected by \citet{Aguado17}, including samples from \citet{Frebel05}, \citet{Sivarani06}, \citet{Frebel06}, \citet{Yong13}, \citet{Allende15}, and \citet{Aguado16}. The values shown for the 11 stars below [Fe/H]$< -4.5$ are plotted separately and their values are taken from \citet{Collet06, Frebel08,Caffau11,Norris12,Bessell15,Hansen15,Frebel15,Bonifacio15,Caffau16, Nordlander17, Bonifacio18a,Aguado18a}; and \citet{Aguado18b}.  All data shown as blue smaller circles are derived using 1D-LTE. Where possible we also show 3D-LTE values, these are shown as larger circles and connected to the 1D-LTE values for the same star by a dotted line. For two stars, this line is vertical, as only 3D corrections for A(C) are available and not for [Fe/H]. The most iron-poor star, SMSS J031300.36-670839.3, has an analysis in 3D-non-LTE for its upper limit of [Fe/H] by \citet{Nordlander17} and this is the value shown here. The most metal-poor star in the literature, SDSS J102915+172927, is labelled by a blue filled circle as well as a small white triangle, HE 0557-4840 is highlighted similarly by a small white square. The horizontal lines give the average A(C) for the various carbon-enhanced star groups as defined by \citet{Spite13}. \label{fig:AC}}
\end{figure}

We note that most metal-poor star known, SDSS J102915+172927, as discovered by \citet{Caffau11}, has similar stellar parameters to \pris, including a very similar colour, ($g-z$) = 0.592. \citet{Caffau12} derive a \teff\ of 5811 $\pm$ 150 K, a microturbulent velocity of 1.5 \kms, and a log(g) of 4.0 $\pm 0.5$ dex (they favoured a larger log(g) rather than a smaller log(g) from the Ca ionisation balance and indeed Gaia DR2 confirms that the star is still on the main-sequence from its larger parallax of 0.734 $\pm$ 0.078, see also \citealt{Bonifacio18c}). The 1D LTE [Fe/H] = $-4.73 \pm 0.13$ is also consistent with the analysis of \pris. The $\alpha$-element abundances in SDSS J102915+172927 are significantly smaller, though. In Figure \ref{fig:compSDSS}, we compare the spectrum that we obtained for Pristine\_221.8781+9.7844 directly with SDSS J102915+172927, the most metal-poor star known as studied by \citet{Caffau11} using exactly the same UVES setup and spectral reduction techniques. Several items stand out when comparing these two spectra. First of all, we see that the Ca II lines are quite comparable in strength (note that in both spectra one can also see features of interstellar Ca that are resolved at this resolution, and that redwards of Ca II H, we see the strong H$\epsilon$ feature). The Mg triplet is stronger in \pris, but the Fe lines are comparable in strength, confirming again an elevation in the $\alpha$-elements in \pris\ compared to SDSS J102915+172927.

Figure \ref{fig:AC} presents the results for our star in 1D LTE and, tentatively, in 3D LTE by straightforwardly applying the corrections of \citet{Caffau12} for SDSS J102915+172927 in  A(C)--[Fe/H] space. Overplotted are a literature sample of extremely metal-poor stars as collected by \citet{Aguado17} based on samples from \citet{Frebel05}, \citet{Sivarani06}, \citet{Frebel06}, \citet{Yong13}, \citet{Allende15}, and \citet{Aguado16}. The 12 stars shown in Figure \ref{fig:AC} below [Fe/H]$< -4.5$ use the measurements from \citet{Collet06, Frebel08,Caffau11,Norris12,Bessell15,Hansen15,Frebel15,Bonifacio15,Caffau16, Nordlander17, Bonifacio18a, Aguado18a, Aguado18b}. Where available, we show both the 1D LTE and 3D LTE measurements as connected symbols of different sizes. For SMSS J031300.36-670839.3, the value of [Fe/H] $<-7.0$ corresponds to the analysis from combined 3D-non-LTE from \citet{Nordlander17}, for all other stars and abundances 3D LTE analyses are shown. 

\begin{figure*}
\centering
\includegraphics[width=\linewidth]{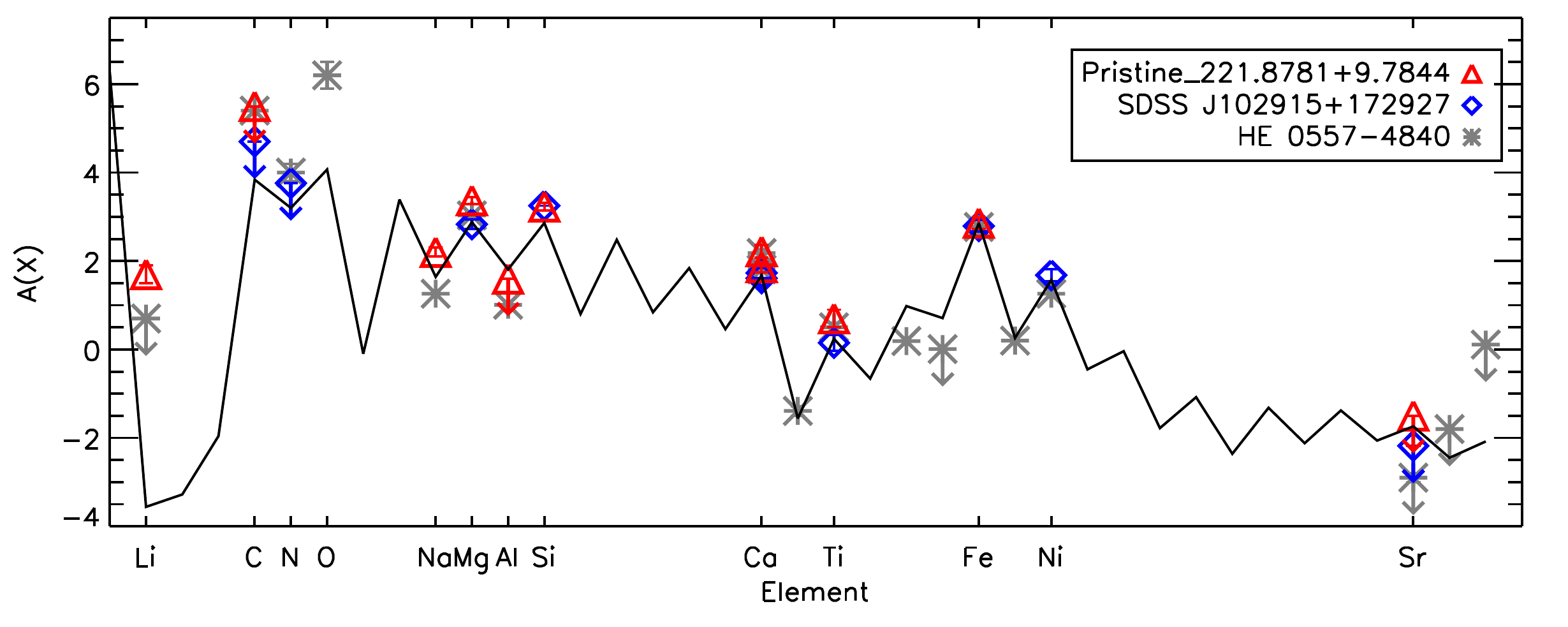}
\caption{A comparison of the 1D LTE abundances for \pris\ (red triangles, data from this work), SDSS J102915+172927 \citep[blue diamonds, data from ][]{Caffau12}, and HE 0557-4840 \citep[grey asterisks, data from][]{Norris07,Norris12}, overplotted by a shifted solar abundance scale with [Fe/H] = $-$4.66 following the solar abundances from \citet{Lodders09} with C and Fe solar abundances taken from \citet{Caffau11b}. \label{fig:abcomp}}
\end{figure*}

It is clear that \pris\ has a very low metal abundance --- not just in [Fe/H], but also in the combination of [Fe/H] and A(C). Because of the relatively large contribution of carbon to the total budget of metals in most stars, this means it is among the most metal-poor stars known. Our upper limit for A(C) in 1D LTE places this star just above the [C/Fe] = +1 line. If we simply adopt the same 3D corrections for carbon and iron from \citet{Caffau12} -- justified by the similarity of both stars in abundance space and stellar parameters -- the upper limit falls instead on the [C/Fe] = +1 dividing line between carbon-enhanced and carbon-normal stars according to the definition of \citet{Beers05}. This would suggest the star is carbon-normal, or even carbon-depleted, according to this definition \citep[although in some other studies a level of +0.7 instead of +1.0 is adopted for carbon-rich stars; see, e.g.,][]{Aoki07}. In any case, \pris\ falls clearly at the low end of the lowest carbon band as defined by \citet{Spite13} (and represented by the horizontal lines in Figure \ref{fig:AC}). Similarly, in the classification of \citet{Yoon16}, it would fall in their Group II (if it would be classified as a carbon-enhanced star). 

Besides its similarity to SDSS J102915+172927, \pris\  is quite close in A(C)--[Fe/H] space to HE 0557-4840 \citep{Norris07,Norris12}. This star has \teff\ = 4900 K and log(g) = 2.2 and is situated on the red giant branch and thus not in the same evolutionary state as either \pris\  or SDSS J102915+172927. It clearly belongs to the ultra metal-poor class of stars with [Fe/H] = $-4.8$ and, above all, it has only a moderate enhancement in the C, N, and O abundances with [C/Fe]= +1.1, [N/Fe] $<$ +0.1, and [O/Fe] = +1.4 \citep[all corrected for 3D effects, see][]{Norris12}. These stars have clearly a lower carbon abundance than the majority of the stars in this range of metallicity. In Figure \ref{fig:abcomp}, the 1D LTE abundance patterns for \pris\, SDSS J102915+172927, and HE 0557-4840 are directly compared.  It is clear that \pris\, and to a lesser extent HE 0557-4840, are a bit more enhanced in $\alpha$-elements than SDSS J102915+172927, but overall the abundance patterns are quite similar for the elements measured in multiple stars.  

\section{Conclusions}\label{sec:discuss}
In this paper, we present the discovery of the ultra metal-poor sub-giant star, Pristine\_221.8781+9.7844, from Ca H\&K narrow-band photometry. We also present an analysis of follow-up medium- and high-resolution spectra using a variety of analysis methods. \pris\  is found to be similar to the most metal-poor star known \citep[SDSS J102915+172927,][]{Caffau11} in terms of stellar parameters, as well as [Fe/H] in standard 1D LTE analysis. A direct comparison of the standard 1D LTE abundances and the spectra (see Figure \ref{fig:compSDSS}) reveals, however, that \pris\  has an [$\alpha$/Fe] ratio of 0.3--0.4 dex, significantly larger than that of SDSS J102915+172927. This is most clearly evident in a stronger Mg triplet feature. Like SDSS J102915+172927, it has no detectable CH features. This leaves open the possibility that this star is carbon-normal, or even carbon-depleted, which would be an anomaly at this extremely low [Fe/H] level. \pris\  also bears a striking resemblance in abundance pattern to HE 0557-4840 \citep{Norris07,Norris12}. Lacking any clear measurement of C, N, and O features in the spectrum of \pris\ and SDSS J102915+172927, it would be premature to argue which star is the most metal-deficient overall, and this is perhaps not the most pressing question at this time. Rather, it is clear that these objects belong to a class of rare, ultra metal-deficient stars that can provide important constraints on cooling and formation of long-lived stars in the low metallicity environment of the early Galaxy. 

\section*{Acknowledgements} 
 
The authors thank Ryan Leaman for insightful discussions and the anonymous referee for careful comments that helped to improve the manuscript. We gratefully thank the CFHT staff for performing the observations in queue mode, for their flexibility in adapting the schedule, and for answering our questions during the data reduction process. We thank Nina Hernitschek for granting us access to the catalogue of Pan-STARRS variability catalogue, allowing a much cleaner target selection within the \textit{Pristine} survey. ES, AA, and KY gratefully acknowledge funding by the Emmy Noether program from the Deutsche Forschungsgemeinschaft (DFG). RAI, NL, NFM, and FS gratefully acknowledge funding from CNRS/INSU through the Programme National Galaxies et Cosmologie and through the CNRS grant PICS07708. FS thanks the Initiative dÕExcellence IdEX from the University of Strasbourg and the Programme Doctoral International PDI for funding his PhD. ES and KY benefited from the International Space Science Institute (ISSI) in Bern, CH, thanks to the funding of the Team ``The Formation and Evolution of the Galactic Halo''.  DA acknowledges the Spanish Ministry of Economy and Competitiveness (MINECO) for the financial support received in the form of a Severo-Ochoa PhD fellowship, within the Severo-Ochoa International Ph.D. Program. DA, CAP, and JIGH also acknowledge the Spanish ministry project MINECO AYA2014-56359-P. JIGH. acknowledges financial support from the Spanish Ministry of Economy and Competitiveness (MINECO) under the 2013 Ram\'{o}n y Cajal program MINECO RYC-2013-14875. HK is financially supported by A. Helmi's VICI grant from the Netherlands Organisation for Scientific Research, NWO. KAV and CLK acknowledge funding from the Discovery Grants and CREATE Program of the National Sciences and Engineering Research Council of Canada. CL thanks the Swiss National Science Foundation for supporting this research through the Ambizione grant PZ00P2\_168065.

Based on observations obtained with MegaPrime/MegaCam, a joint project of CFHT and CEA/DAPNIA, at the Canada-France-Hawaii Telescope (CFHT) which is operated by the National Research Council (NRC) of Canada, the Institut National des Science de l'Univers of the Centre National de la Recherche Scientifique (CNRS) of France, and the University of Hawaii. The observations at the Canada-France-Hawaii Telescope were performed with care and respect from the summit of Maunakea which is a significant cultural and historic site. 

The William Herschel Telescope is operated on the island of La Palma by the Isaac Newton Group of Telescopes in the Spanish Observatorio del Roque de los Muchachos of the Instituto de Astrof'sica de Canarias.

Funding for the Sloan Digital Sky Survey IV has been provided by the Alfred P. Sloan Foundation, the U.S. Department of Energy Office of Science, and the Participating Institutions. SDSS-IV acknowledges support and resources from the Center for High-Performance Computing at the University of Utah. The SDSS web site is www.sdss.org. SDSS-IV is managed by the Astrophysical Research Consortium for the Participating Institutions of the SDSS Collaboration including the Brazilian Participation Group, the Carnegie Institution for Science, Carnegie Mellon University, the Chilean Participation Group, the French Participation Group, Harvard-Smithsonian Center for Astrophysics, Instituto de Astrof\'isica de Canarias, The Johns Hopkins University, Kavli Institute for the Physics and Mathematics of the Universe (IPMU) / University of Tokyo, Lawrence Berkeley National Laboratory, Leibniz Institut f\"ur Astrophysik Potsdam (AIP), Max-Planck-Institut f\"ur Astronomie (MPIA Heidelberg), Max-Planck-Institut f\"ur Astrophysik (MPA Garching), Max-Planck-Institut f\"ur Extraterrestrische Physik (MPE), National Astronomical Observatories of China, New Mexico State University, New York University, University of Notre Dame, Observat\'ario Nacional / MCTI, The Ohio State University, Pennsylvania State University, Shanghai Astronomical Observatory, United Kingdom Participation Group,Universidad Nacional Aut\'onoma de M\'exico, University of Arizona, University of Colorado Boulder, University of Oxford, University of Portsmouth, University of Utah, University of Virginia, University of Washington, University of Wisconsin, Vanderbilt University, and Yale University.

This work has made use of data from the European Space Agency (ESA) mission {\it Gaia} (\url{https://www.cosmos.esa.int/gaia}), processed by the {\it Gaia}
Data Processing and Analysis Consortium (DPAC, \url{https://www.cosmos.esa.int/web/gaia/dpac/consortium}). Funding for the DPAC has been provided by national institutions, in particular the institutions participating in the {\it Gaia} Multilateral Agreement.

\appendix

\section{Line list}

Table \ref{tab:linelist} presents the used line list for Fe lines; Table \ref{tab:linelist2} lists all lines used in the analyses of other elements. Measurement uncertainties for the equivalent width methods are not presented, as the total uncertainties are instead dominated by the continuum placement. This is the main source of the systematic discrepancy between the two equivalent width methods (methods 3 and 4). Despite these discrepancies in measurements, the resulting abundance determinations are compatible within their uncertainties (as illustrated in Section \ref{sec:abu}, with the exception of Ca and Ti that show 0.1 dex difference).

\begin{table*}
\centering
\begin{tabular}{ccrrcccc} 
\hline 
Wavelength & Ion & $\chi$ & log(gf) & used & used & EW (m$\AA$) & EW (m$\AA$) \\
($\AA$) & & & & method1& method2 & method3 & method4\\
\hline
\hline
5328.531 & \ion{Fe}{i} &  1.56 & $-$1.85 & X & -- & -- & -- \\
5269.537 & \ion{Fe}{i} & 0.86 & $-$1.33 & X & X & -- & 14 \\
4415.122 & \ion{Fe}{i} &  1.61 & $-$0.62 & X & -- & 10 & -- \\
4404.750 & \ion{Fe}{i} & 1.56 & $-$0.15 & X & X & 19 & 22 \\
4383.545 & \ion{Fe}{i} & 1.48 & 0.21 & X & X & 32 & 38 \\
4325.762 & \ion{Fe}{i} & 1.61 & 0.01 & X & X & 13 & 15 \\
4307.910 & \ion{Fe}{i} & 1.56 & $-$0.07 & -- & X & -- & 15 \\
4294.140 & \ion{Fe}{i} & 1.49 & $-$1.11 & --& -- & -- & 6 \\
4271.761 & \ion{Fe}{i} & 1.48 & $-$0.17 & X & X & 18 & 22 \\
4260.474 & \ion{Fe}{i} & 2.40 & 0.08 & X & -- & 11 & 10 \\
4202.029 & \ion{Fe}{i} & 1.48 & $-$0.69 & X & -- & -- & -- \\
4143.868 & \ion{Fe}{i} & 1.56 & $-$0.51 & X & X & -- & 7 \\
4071.738 & \ion{Fe}{i} & 1.61 & $-$0.01 & X & X & 19 & 23 \\
4067.978 & \ion{Fe}{i} &  3.21& $-$0.53 & X & -- & -- & -- \\
4063.594 & \ion{Fe}{i} & 1.56 & 0.06 & X & X & 16 & 23 \\
4045.812 & \ion{Fe}{i} & 1.48 & 0.28 & X & -- & 30 & 29 \\
4005.242 & \ion{Fe}{i} & 1.56 & $-$0.58 & X & -- & -- & 12 \\
3930.297 & \ion{Fe}{i} & 0.09 & $-$1.49 & -- & X & -- & 29 \\
3927.921 & \ion{Fe}{i} & 0.11 & $-$1.52 & -- & X & -- & 29 \\
3922.912 & \ion{Fe}{i} & 0.05 & $-$1.63 & X & X & 17 & 19 \\
3920.258 & \ion{Fe}{i} & 0.12 & $-$1.73 & X & X & 10 & 15 \\
3906.479 & \ion{Fe}{i} & 0.11 & $-$2.21 & -- & X & -- & -- \\
3902.945 & \ion{Fe}{i} & 1.56 &	$-$0.44 & -- & -- & 11 & -- \\
3899.707 & \ion{Fe}{i} & 0.09 & $-$1.51 & X & X & 23 & 33 \\
3895.656 & \ion{Fe}{i} & 0.11 & $-$1.67 & X & X & 23 & 24 \\
3887.048 & \ion{Fe}{i} & 0.91 & $-$1.09 & X & --  & -- & -- \\
3886.282 & \ion{Fe}{i} & 0.05 & $-$1.05 & X & X & 45 & 52 \\
3878.573 & \ion{Fe}{i} & 0.09 & $-$1.38 & -- & X & 22 & 27 \\
3878.018 & \ion{Fe}{i} & 0.96 & $-$0.90 & -- & X & 17 & -- \\
3865.523 & \ion{Fe}{i} & 4.14 & $-$0.93 & X & X & -- & -- \\
3859.912 & \ion{Fe}{i} & 0.00 & $-$0.70 & X & -- & 56 & 62 \\
3859.213 & \ion{Fe}{i} & 2.40 & $-$0.68 & -- & X & -- & -- \\
3856.372 & \ion{Fe}{i} & 0.05 & $-$1.28 & X & X & 35 & 36 \\
3849.967 & \ion{Fe}{i} & 1.01 & $-$0.86 & X & -- & 16 & -- \\
3841.048 & \ion{Fe}{i} & 1. 61 & $-$0.04 & X & -- & -- & -- \\
3840.438 & \ion{Fe}{i} & 0.99  & $-$0.50 & X &  -- & -- & -- \\
3834.222 & \ion{Fe}{i} & 0.96  & $-$0.27 & -- & X & -- & -- \\
3827.823 & \ion{Fe}{i} & 1.56  &  0.09 & X & -- & 15 & -- \\
3825.881 & \ion{Fe}{i} & 0.91 & $-$0.02 & X & X & 33 & 43 \\
3824.444 & \ion{Fe}{i} & 0.00 & $-$1.34 & X & X & 30 & 36  \\
3820.425 & \ion{Fe}{i} & 0.86 & 0.16 & X & X & 48 & 59 \\
3815.840 & \ion{Fe}{i} & 1.48 & 0.24 & X & -- & 30 & 35 \\
3812.964 & \ion{Fe}{i} & 0.96 & $-$1.05 & -- & X & -- & -- \\
3767.192 & \ion{Fe}{i} &  1.01 &  $-$0.38 & X & X & -- & -- \\
3763.789 & \ion{Fe}{i} & 0.99 & $-$0.22 & X & X & 28 & 35 \\
3758.233 & \ion{Fe}{i} & 0.96 & $-$0.01 & -- & X &  37 & 38 \\
3745.561 & \ion{Fe}{i} & 0.09 & $-$0.77 & -- & X & -- & 50 \\
3745.899 & \ion{Fe}{i} & 0.12 & $-$1.34 & -- & -- & -- & 28 \\
3748.262 & \ion{Fe}{i} & 0.11 & $-$1.01 & -- & -- & -- & 41 \\
3737.131 & \ion{Fe}{i} & 0.05 & $-$0.57 & -- & X & -- & 56 \\
3734.884 & \ion{Fe}{i} & 0.86 &	0.33 & -- & X &   & -- \\
3727.619 & \ion{Fe}{i} & 0.96  &  $-$0.61 & X & -- &  17 & -- \\
3722.563 & \ion{Fe}{i} & 0.09 & $-$1.28 & -- & X & -- & -- \\
3719.935 & \ion{Fe}{i} & 0.00 & $-$0.42 & -- & -- & 73 & 79 \\
3705.566 & \ion{Fe}{i} & 0.05 & $-$1.32 & -- &  -- & -- & 32 \\
3709.246 & \ion{Fe}{i} & 0.91 & $-$0.62 & -- & -- & -- & 29 \\
\hline
\end{tabular}
\caption{Line list of Fe lines used in the analysis.\label{tab:linelist}}
\end{table*}

\begin{table*}
\centering
\begin{tabular}{ccrrcccc} 
\hline 
Wavelength & Ion & $\chi$ & log(gf) & used & used & EW (m$\AA$) & EW (m$\AA$) \\
($\AA$) & &  & & method1& method2 & method3 & method4\\
\hline
\hline
5895.924 & \ion{Na}{i} & 0.00 & $-$0.19 & X & -- & -- & 13 \\
5889.970 & \ion{Na}{i} & 0.00 & 0.11 & X & -- &  25 & 21 \\
5183.604 & \ion{Mg}{i} & 2.72 & $-$0.16 & X & X & 39 & 37 \\
5172.700 & \ion{Mg}{i} & 2.71 & $-$0.38 & X & X & 30 & 30 \\
5167.31 & \ion{Mg}{i} & 2.71 & $-$1.03 & X & X & -- &12 \\
3838.292 & \ion{Mg}{i} & 2.72 & 0.42 & X & -- & -- & 64 \\
3832.304 & \ion{Mg}{i} & 2.71 & 0.150 & X & -- & 49 & 50 \\
3829.355 & \ion{Mg}{i} & 2.71 & $-$0.23 & X & -- & 24 & 31 \\
3905.523 & \ion{Si}{i} & 1.91 & $-$1.09 & X & X & 23 & 22 \\
4226.728 & \ion{Ca}{i} & 0.00 & 0.24 & X & X & 32 & 37 \\
3736.902& \ion{Ca}{ii} & 3.15 & $-$0.15 & X & -- & -- & -- \\
3933.663& \ion{Ca}{ii} & 0.00 & 0.13 & X & -- & -- & -- \\
3968.470& \ion{Ca}{ii} & 0.00 & $-$0.17 & X & -- & -- & -- \\
3759.292 & \ion{Ti}{ii} & 0.61 & 0.28 & X & -- & 32 & 38 \\
3761.321 & \ion{Ti}{ii} & 0.57 & 0.18 & X & X & 28 & 35 \\
4077.710 & \ion{Sr}{ii} & 0.00 & 0.17 & X & --& 12 & 7 \\
\hline
\end{tabular}
\caption{Line list of elements other than Fe used in the analysis.\label{tab:linelist2}}
\end{table*}

%%%%%%%%%%%%%%%%%%%%%%%%%%%%%%%%%%%%%%%%%%%%%%%%%%



\bibliographystyle{mn2e}
\bibliography{references} % if your bibtex file is called example.bib

% Don't change these lines
\bsp	% typesetting comment
\label{lastpage}
\end{document}